\documentclass{aa}  

\usepackage{graphicx}
\usepackage{orcidlink}
\usepackage{txfonts}
\usepackage{tabularx}
\usepackage{booktabs}
\usepackage{placeins}
   \newcommand{\unit}[1]{\ensuremath{\, \mathrm{#1}}}
   \newcommand{\arepo}{{\sc arepo}}          % code names
   \newcommand{\polaris}{{\sc polaris}}          % code names
\begin{document}

   \title{Feasibility of detecting shadows in disks induced by infall}

   \subtitle{}

   \author{A. Krieger\inst{1}  \orcidlink{0000-0002-3639-2435}
          \and
          M. Kuffmeier
          \inst{2}\fnmsep \orcidlink{0000-0002-6338-3577}
          \and
          S. Reissl\inst{3} \orcidlink{0000-0001-5222-9139}
          \and
          C. P. Dullemond\inst{3} \orcidlink{0000-0002-7078-5910}
          \and
          C. Ginski\inst{4} \orcidlink{0000-0002-4438-1971}
          \and
          S.~Wolf \inst{1} \orcidlink{0000-0001-7841-3452}
          }

   \institute{Institut für Theoretische Physik und Astrophysik, Christian-Albrechts-Universität zu Kiel, Leibnizstra{\ss}e 15, 24118 Kiel, Germany \\  \email{akrieger@astrophysik.uni-kiel.de}            
               \and
   Niels Bohr Institute, University of Copenhagen, {\O}ster Voldgade 5, DK-1350 Copenhagen, Denmark\\     \email{kueffmeier@nbi.ku.dk}          
               \and
   Zentrum für Astronomie der Universität Heidelberg, Institut für Theoretische Astrophysik, Albert-Ueberle-Stra{\ss}e 2, 69120 Heidelberg
             \and
   School of Natural Sciences, Center for Astronomy, University of Galway, Galway, H91 CF50, Ireland             }

   \date{Received \today}

% \abstract{}{}{}{}{} 
% 5 {} token are mandatory

  \abstract
  % context heading (optional)
  % aims heading (mandatory)
  % methods heading (mandatory)
  % results heading (mandatory)
  % conclusions heading (optional), leave it empty if necessary 
    {Observations performed with high-resolution imaging techniques revealed the existence of shadows in circumstellar disks that can be explained by the misalignment of an inner with respect to an outer disk. The cause of misalignment, however, is still debated. In this study, we investigate the feasibility of observing shadows induced by one prominent scenario that may lead to misalignment, which involves the late infall of material onto a protostellar system. In particular, we use previously performed hydrodynamical simulations of such events, and generate flux maps in the visible, near-infrared, submillimeter, and millimeter wavelength range using Monte Carlo radiative transfer. Based on that, we derive synthetic observations of these systems performed with the instruments SPHERE/VLT and ALMA, which we use as a basis for our subsequent analysis. We find that near-infrared observations with SPHERE are particularly well suited for detecting shadows via direct imaging alongside other features such as gaps, arcs, and streamers. On the contrary, performing a shadow detection based on reconstructed ALMA observations is very challenging due to the high sensitivity that is required for this task. Thus, in cases that allow for a detection, sophisticated analyses may be needed, for instance by the utilization of carefully constructed azimuthal profiles, aiding the search for potentially shallow shadows. Lastly, we conclude that late infall-induced disk misalignment offers a plausible explanation for the emergence of shadows that are observed in various systems. 
   }


   \maketitle
%
%-------------------------------------------------------------------
\section{Introduction}
Circumstellar disks are the birth locations of planets and form as a by-product of the star-formation process.
The capability of performing high-contrast and high-sensitivity observations of these disks with modern telescopes allows us to constrain their properties, in particular their spatial structures. 
Especially, observations of scattered light revealed shadow features in various systems, including PDS~66 \citep{2016ApJ...818L..15W}, HD~142527 \citep{2017AJ....154...33A}, HD~100453 \citep{Benisty2017}, HD~169142 \citep{2018MNRAS.474.5105B}, RXJ1604.3-2130 \citep{2018ApJ...868...85P}, HD~139614 \citep{2020A&A...635A.121M}, HD~34700 \citep{2020ApJ...900..135U}, SU~Aur \citep{Ginski2021}, and TW~Hya \citep{2023ApJ...948...36D}. Moreover, shadows were detected for dust continuum observations of MWC~758 \citep{2018ApJ...853..162B}, for dust continuum and $^{12}$CO J = 2–1 line emission of RXJ1604.3-2130 \citep{2023A&A...670L...1S}, as well as for observations of the $^{12}$CO J = 2–1 line of TW Hya \citep{Teague+2022}.
Recent studies suggest, that the presence of such features in scattered light observations of disks performed with the Spectro-Polarimetric High contrast imager for Exoplanets REsearch \citep[SPHERE;][]{2019A&A...631A.155B} at $\sim 1\,\mu$m (H, R', I', J, and K bands) can in many cases be best explained by a configuration, where an inner disk is misaligned with respect to an outer disk \citep{Avenhaus2014,Marino2015,Benisty2017,Benisty2018,Casassus2018,Ginski2021}.
In such a misaligned configuration, the illumination of selected parts of the outer disk by the central star is prevented by the inner disk.  
While there is consensus that misalignment between inner and outer disk is a good explanation for the presence of the observed shadows, there is an ongoing debate about the origin of misalignment \citep[see PPVII reviews][]{Benisty2022,Pinte2022}. 
Broadly speaking, one can distinguish between two main scenarios.
The first scenario suggests that the primordial disk breaks apart and subsequently the inner and outer disk become misaligned to each other. 
The break-up and misalignment can be induced by a perturber such as an embedded massive planet, brown dwarf, or binary companion that may be located inside \citep{Nixon2013,2017MNRAS.469.2834O,Nealon2018,Zhu2019} or outside the disk \citep{Dogan2015}. 
Also, a combination of an inner perturber, such as a planet, and an external perturber, such as a wide-orbit binary companion, is discussed to explain this phenomenon \citep{Gonzalez2020,Nealon2020b}. 
In the other scenario, the outer disk forms from material with a different orientation of net angular momentum than the inner disk that has already formed earlier \citep{Thiess2011,Kuffmeier2021}.
This idea is in line with the possibility of infall onto protostellar systems that form in turbulent birth environments of Giant Molecular clouds as seen in models \citep{Padoan2014,Kuffmeier2017,Kuffmeier2023,Bate2018} and as observed in star-disk systems that are associated with streamers \citep[e.g.,][]{LeGouellec+2019,Pineda2020,Valdivia-Mena+2022,Valdivia-Mena+2023}, \citep[see also][for a recent review of star formation including constraints from asymmetric infall via streamers]{Pineda+2023}.

Motivated by these predictions of infall in models, and considering observations showing extended arm structures around presumably more evolved Class II Young Stellar Objects such as AB Aur \citep{Grady1999}, SU Aur \citep{Akiyama+2019,Ginski2021}, GM Aur, \citep{Huang+2021}, Elias 2-27 \citep{Paneque-Carreno+2021}, DR Tau \citep{Mesa+2022,Huang+2023}, RU Lup \citep{Huang+2020}, DO Tau \citep{Huang+2022} as well as strong indications of late infall from statistics of reflection nebulae \citep{Gupta+2023}, we explore a scenario that links the occurrence of late infall to the observation of shadows. In this study, we investigate the feasibility of detecting observable shadow features induced by misaligned disks that have formed as a result of late infall. 
For that purpose, we present and analyze maps of scattered and thermally emitted light from the dust phase that are based on snapshots of previously conducted hydrodynamical simulations of star-disk encounters with an infalling cloudlet \citep[][]{Kuffmeier2021}. These maps are subsequently post-processed to generate synthetic observations at different observing wavelengths. This allows us to test whether infall is a plausible explanation for the shadow features found in scattered light, thermal continuum emission, and line emission observations of real systems. 

\section{Methods}
In the following, we describe the procedure we applied to generate synthetic observations, i.e., the post-processing of results of radiative transfer simulations which are based on previously performed hydrodynamical simulations.
The underlying hydrodynamical simulations were carried out with the moving-mesh code \arepo \footnote{\url{https://arepo-code.org/}}\citep{Springel2010,Pakmor2016}. 
The setup of the hydrodynamical simulations are presented in detail in a recent paper \citep[][]{Kuffmeier2021} and we only briefly summarize the main features of the underlying models. 
In the hydrodynamical runs, it was investigated how the gravitational potential of a star affects the dynamics of a cloudlet of gas that passes by the star.  
As a consequence of angular momentum conservation, such an encounter event leads to the formation of a new disk if the cloudlet has a non-zero impact parameter $b$.
The cloudlet was initialized at location 
\begin{equation}
    \mathbf{r}_{\rm init, cloudlet} = 
    \begin{pmatrix} 
       x_{\rm cloudlet} \\
       y_{\rm cloudlet} \\
       z_{\rm cloudlet} 
    \end{pmatrix} = 
    \begin{pmatrix} 
       -3.22\, R_{\rm cloudlet} \\
       -b \cos{\alpha}  \\
       -b \sin{\alpha} 
    \end{pmatrix}, 
\end{equation}
where $\alpha$ is the infall angle of the cloudlet measured with respect to the $xy$-plane of the coordinate system.
The cloudlet radius $R_{\rm cloudlet}=887\,$au and the impact parameter $b = 1774\,$au correspond to $0.4\,b_{\rm crit}$ and $0.8\,b_{\rm crit}$, respectively. Here, $b_{\rm crit}$ describes the impact parameter at which a test particle that encounters a $2.5\,{\rm M}_{\odot}$ star with velocity $v_{\rm i}$ would be deflected by $90^{\circ}$ \citep[see Sect.~2 in ][]{Dullemond2019}.
The mass of the cloudlet was set to 
\begin{equation}
    M_{\rm cloudlet}(R_{\rm cloudlet}) = 0.01 {\rm M}_{\odot} \left( \frac{R_{\rm cloudlet}}{5000 \unit{au}}\right)^{2.3}.
    \label{KleHen}
\end{equation}
It was initialized with a uniform density distribution $\rho_{\rm cloudlet}$, and the density of the background gas was set to $\rho_{\rm bg}=\frac{1}{800}\rho_{\rm cloudlet}$. 
Furthermore, we enforced turbulent motions within the cloudlet \citep[for more details see][]{Kuffmeier2020,Kuffmeier2021}.
We assumed an isothermal temperature for the gas of 10 K. 
The cloudlet approaches the central star (modeled as a point source with mass $M_{*} = 2.5 \unit{M}_{\odot}$) with a velocity of 
\begin{equation}
    \mathbf{v}_{\rm cloudlet} = 
    \begin{pmatrix} 
       v_{\rm x, cloudlet} \\
       v_{\rm y, cloudlet} \\
       v_{\rm z, cloudlet} 
    \end{pmatrix} = 
    \begin{pmatrix} 
       v_{\rm i} = 10^5 \unit{cm}\unit{s}^{-1} \\
       0  \\
       0 
    \end{pmatrix},
\end{equation} 
while the background gas is at rest with respect to the central star.

The star hosts a circumstellar disk with an initial density profile of $\Sigma = \Sigma_0 \left(\frac{r}{\rm au}\right)^{-p}$ between $20 \unit{au}<r<50 \unit{au}$. 
Inside and outside the disk, this profile was tapered off to the background density using a logistic function. 
In this paper, we focus on post-processing the previously performed runs with $\Sigma_0=170 \unit{g}\unit{cm}^{-2}$ and $p=1.5$. 
We considered infalling angles $\alpha$ of $35^{\circ}$, $60^{\circ}$, and $90^{\circ}$ \citep[i.e., runs 3, 4, 5, 10, and 11 in][]{Kuffmeier2021}. 
For angles of $35^{\circ}$ and $60^{\circ}$ the inner disk was either set up with prograde (in runs 3 and 4, respectively) or retrograde rotation (in runs 10 and 11, respectively) with regard to the rotation of the newly forming outer disk. 
In this study, we mainly show and discuss results of the case of perpendicular infall \citep[i.e., run 5 in][column density plots are shown in Fig. \ref{fig:column_density}]{Kuffmeier2021} with an inclination angle of the infalling cloudlet of $\alpha=90^{\circ}$.

\subsection{Radiative transfer post-processing}
\label{sec:radiative_transfer_pos_processing}
In order to simulate synthetic observations of scattered light and dust emission, we utilize the publicly available Monte Carlo (MC) radiative transfer (RT) code \polaris\footnote{ \url{https://portia.astrophysik.uni-kiel.de/polaris/}} \citep[see][]{Reissl2016,Reissl2019,Reissl2020}. \polaris\, uses a Voronoi grid similar to the native grid of the \arepo\, code for performing RT simulations, hence, no re-gridding of the HD data is required. The dust is modeled assuming spherical grains of distinct radii $a_{\mathrm{eff}}$ and with optical properties corresponding to a mixture of $37.5\ \%$ graphite and $62.5\ \%$ (astro)silicate \citep{DraineLee1984,LaorDraine1993,Draine2003}. The grain size distribution follows a power-law ${ {\propto}\, a_{\mathrm{eff}}^{-3.5} }$  typical for the ISM \citep[for details, see e.g.][]{Mathis1977,LiDraine2001}. However, for the range of grain radii we assume a considerable grain growth within the disks leading to a range of ${ a_{\mathrm{eff}} \in [5\, \mathrm{nm}, 10\, \mu\mathrm{m}] }$. For the spatial distribution of the dust we utilize a constant dust mass to gas mass ratio of $1\ \%$.

In a first post-processing step, we calculate the underlying dust temperature, assuming an equilibrium of absorbed and re-emitted radiation for each grain size. For the radiation field, we consider the spectral emission of the central star as well as an interstellar radiation field (ISRF) typical for our Galactic neighborhood \cite[][]{Mathis1983}. In a second step, we perform monochromatic MC dust scattering simulations as well as non-probabilistic RT simulations to generate thermal dust emission maps utilizing a ray-tracing algorithm, assuming either of three inclination angles $i\in \left\{0^\circ,\,45^\circ,\,90^\circ\right\}$ and observing wavelengths $\lambda= 0.6263,\, 1.245,\, 2.182,\, 850,$ and $1300\,\mu$m.

\section{Results and discussion}

    In the following, we assess the feasibility of detecting shadows cast onto the outer disk at different observing wavelengths based on synthetic scattering and dust emission maps. In Sect.~\ref{sec:sphere}, we show the results of simulated observations in the visible and near-infrared (VIS/NIR) wavelength range performed with SPHERE, which provides a resolution that is high enough to regularly resolve small-scale structures in protoplanetary disks. In this wavelength range, the observed flux is dominated by stellar radiation that is scattered off the photospheres of the disks. In this case, shadows emerge due to blocked irradiation of the outer disk along the direction in which both disk mid-planes align, leading to a potentially observable contrast between the irradiated and shadowed sections of the outer disk. This effect is for the most part geometrical, and the feasibility of a detection primarily depends on the flux level and emerging contrast. 

    Due to the reduction of heating irradiation, the shadowed regions cool down, resulting in a lowering of observed flux in the submillimeter and millimeter (submm/mm) wavelength range and the formation of apparent shadows, which can also lead to observable arc-like features \cite[see for instance][]{2015ApJ...812..126C}. However, for their emergence, the shielded regions in the outer disk need to cool down sufficiently fast before the relative orientation of the disks changes and the shadowed region shifts. The cooling and heating timescales and their potential effects on observations are the topic of Sect.~\ref{sec:cooling_and_heating}. 
    
    Lastly, in Sect.~\ref{sec:alma}, we simulate observations performed with the Atacama Large Millimeter/submillimeter Array \citep[ALMA;][]{2015ApJ...808L...1A} in the submm/mm wavelength range. In this case, the inner and outer disks are almost optically thin and corresponding flux maps are for the most part determined by the spatial temperature and density distributions of both disks. A successful detection of the emerging shadows in this wavelength range therefore likewise depends on the observed contrast between the shadowed and the adjacent illuminated regions and the total flux of the system. 
    
    % introduction reference star
    Since realistic detection limits for both instruments generally depend on various properties of the observed system, we use HL Tau \citep[RA: 4h31min38s, Dec: +18$^\circ$13'57$"$, J2000,][]{2011ApJ...741....3K} as a proxy, a system composed of a T Tauri star hosting a protoplanetary disk located in the close-by Taurus star-formation region at a distance of ${\sim}140\,$pc \citep{2004AJ....127.1029R}. Its spectral type is K5 \citep{2004ApJ...616..998W}, and its flux values in the J and V bands are 10.624\,mag \citep{2003yCat.2246....0C} and 14.49\,mag \citep{2012yCat.1322....0Z}, respectively. In the context of this study, using HL Tau as a proxy rather than a more evolved, brighter star yields comparably low, and thus conservative, detectable contrast value estimates. Moreover, it is worth mentioning that we will assume different distances to the reference star throughout this paper, which may differ from the actual distance to HL Tau. In particular, we assume a generic value of $d=140\,$pc, roughly corresponding to the distance to the Taurus star-forming region, which allows us to compare our results to observations of the SU Aur system. Additionally, we investigate the feasibility of detecting shadows in more distant systems and assume, for that purpose, a distance of 400\,pc. 

\subsection{SPHERE observations}
\label{sec:sphere}
% https://cdn.sci.news/images/enlarge8/image_9380e-SU-Aur.jpg
% https://ui.adsabs.harvard.edu/abs/2021ApJ...908L..25G/abstract

    \begin{figure}
        \centering
        \includegraphics[width=\hsize]{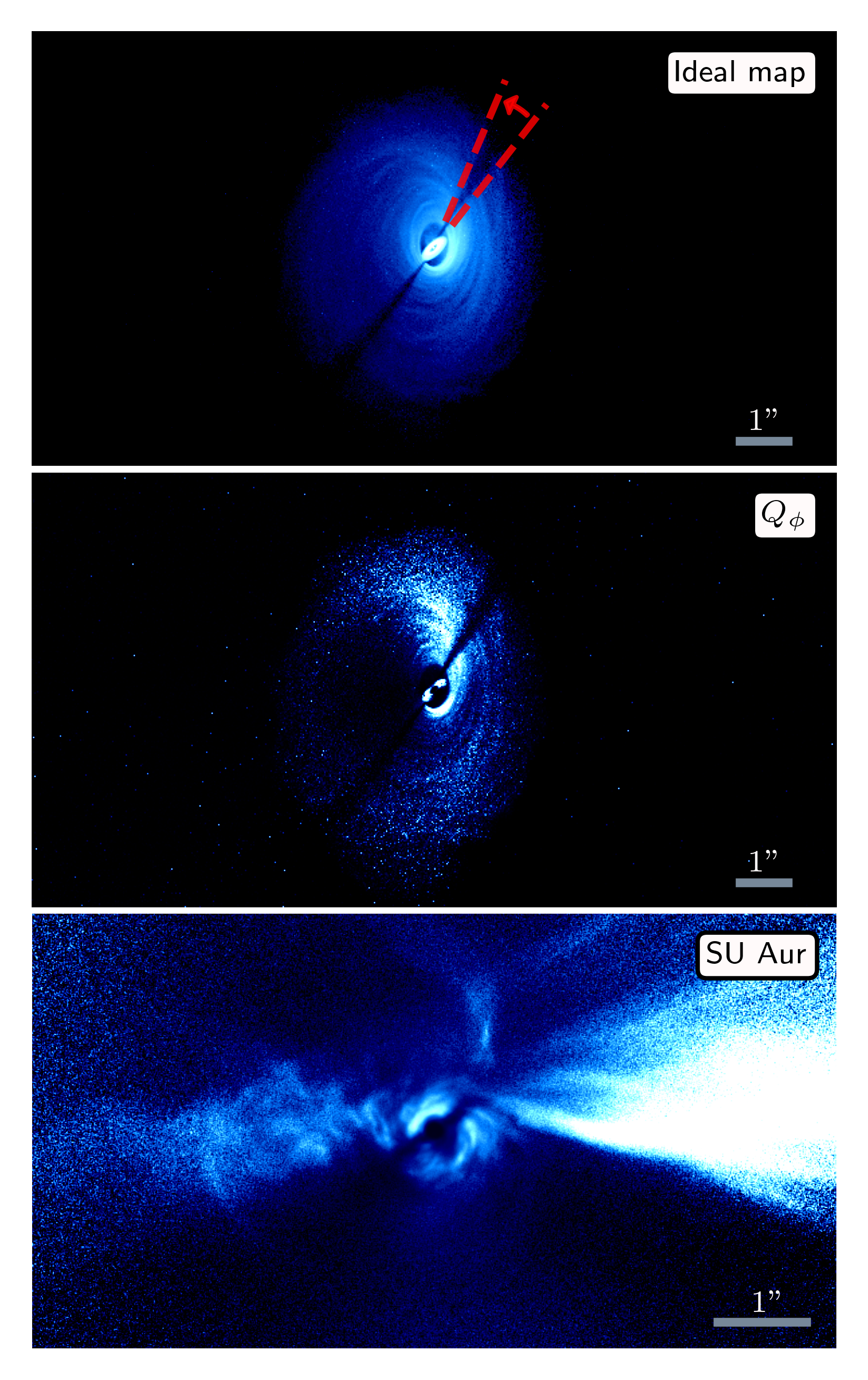} 
            \caption{Top: Ideal flux map at $1.245\,\mu$m for run 5 at $150 \,$kyr assuming a distance of $140\,$pc. Red lines highlight the shadow feature and its opening angle in the upper right quadrant of the image. Mid: Corresponding ideal $Q_\phi$ map. For illustrative purposes, $r^2$-normalization is applied to the image to better highlight the regions far from the central star. Bottom: SPHERE/IRDIS H band $Q_\phi$ observation of the SU Aur system \citep{Ginski2021}.}
            \label{fig:SU_Aur}
    \end{figure}

    To assess the feasibility of detecting shadows in the outer disk, we generate synthetic observations for the instruments IRDIS and ZIMPOL. Both instruments are installed on SPHERE at the telescope UT3 of the VLT with a diameter of $D=8.2\,$m. Furthermore, we assume the usage of broadband filters and coronagraphs to enable high-contrast observations\footnote{SPHERE User Manual of Period 111, Phase 2:\\\url{https://www.eso.org/sci/facilities/paranal/instruments/sphere/doc/VLT-MAN-SPH-14690-0430_P111_dec_2022_zwa.pdf}}. For the instrument ZIMPOL in the VIS wavelength range, we simulate observations performed in the mode ZIMPOL\_I using the broad band filter R\_PRIM with a central wavelength of $\lambda=0.6263\,\mu$m and the recommended classical Lyot coronagraph V\_CLC\_M\_WF with an inner working angle (IWA) of $0.155\,$mas. For the simulated IRDIS observations in the NIR wavelength range, we assume the usage of the classical imaging mode, either using the broad band filter BB\_J with a central wavelength of $\lambda=1.245\,\mu$m or the broad band filter BB\_Ks with a central wavelength of $\lambda=2.182\,\mu$m. For these synthetic observations, we furthermore assume the utilization of the apodized Lyot coronagraphs N\_ALC\_YJH\_S (${\rm IWA}=0.15\,$mas) or N\_ALC\_Ks (${\rm IWA}=0.2\,$mas), respectively. For illustration, Fig.~\ref{fig:SU_Aur} shows an ideal flux map of a simulated system for an observing wavelength of $1.245\,\mu$m in the top plot. 

    On the basis of simulated ideal polarimetric observations of the Stokes parameters $Q$ and $U$ we furthermore create maps of the azimuthal Stokes parameters \citep{2020A&A...633A..63D},
    \begin{eqnarray}
        Q_\phi = -Q \cos{2\phi} - U \sin{2\phi},\\
        U_\phi = +Q \sin{2\phi} - U \cos{2\phi},
    \end{eqnarray}
    where $\phi$ describes the azimuthal coordinate, which for any cartesian pixel at position $(x,\,y)$ is given by
    \begin{equation}
        \phi = \arctan\left(\frac{x-x_*}{y-y_*} \right),
        \label{eq:phi_coordinate}
    \end{equation}
    with $(x_*,\,y_*)$ as the position of the star.  
    In comparison to Eq. \eqref{eq:phi_coordinate}, \citet{2020A&A...633A..63D} used an additional offset angle $\phi_0$, which is not required in our simulations. This quantity was introduced to account for the rotation of the derotator, that was used over the course of multiple observational cycles, which otherwise would have led to a likewise rotation of apparent polarization vectors. Exemplarily, results for an ideal $Q_\phi$ map are shown in the middle plot of Fig.~\ref{fig:SU_Aur}. 
    For comparison purposes, the bottom plot shows the results of observations of the SU Aur system, which were performed with SPHERE/IRDIS in the H band \citep{Ginski2021}. Each of these plots shows, alongside extended arm features, i.e., streamers, two strikingly dark regions extending from their respective centers outward in almost diametrically opposed directions. In order to assess the feasibility of observing such shadows, synthetic observations are constructed on the basis of these ideal flux maps.

    % generate synthetic observation
    A synthetic observation at a given observing wavelength is generated by convolving the corresponding ideal flux map with a gaussian beam using a wavelength-dependent full width at half maximum (FWHM) of $1.22\,\lambda/D$. Other instrument specific properties that may affect the observation have been neglected for this analysis. Subsequently, for the simulated flux of each pixel ($f_{\rm px}$) the corresponding contrast $C_{\rm px} = -2.5\log_{10}(f_{\rm px}/f_{\rm max})$ is calculated, where $f_{\rm max}$ is the maximum pixel value in each map, located at the projected position of the star on the map. Lastly, the central region is masked according to the size of the IWA.

    % detectability / observing criterion
    The feasibility of SPHERE to detect faint sources is limited by the contrast between its detected flux and the flux of the star. The detection limit additionally depends on the specifics of the instrument, in particular, the filter and coronagraph, and the distance between the source and the star. In order to estimate realistic detection limits, we make use of the SPHERE ESO exposure time calculator (ETC)\footnote{\url{https://www.eso.org/observing/etc/}}. Thereby obtained $5\sigma$ performance curves provide the maximum contrast values, i.e., the detection limits, that allow for a detection as functions of the distance to the star. If the contrast between the source and the star exceeds this limit, it is too dim and can thus not be detected. Since these limits generally depend on the celestial coordinates of the observed system as well as on its specific properties, we use properties of the reference star to obtain generic estimates. Furthermore, we assume that the observations are performed using the pupil-stabilized mode, each with an exposure time of 3600\,s as well as a DIT of 8\,s for IRDIS  or 10\,s for ZIMPOL. A list of selected contrast detection limits for the considered three wavelengths is presented in Table \ref{tab:detection_limits}. We note, that particularly for faint disks readout noise can play a critical role, resulting in a need for higher DIT values. We confirmed in a test, though, that the effect of its increase to 64\,s (50\,s) for IRDIS (ZIMOL) had no substantial impact on the qualitative validity of our results. To further ensure the quality of our results, we conducted an additional test to confirm that even a 5\,mag artificial increase of the brightness of the star also did not qualitatively affect our results.
    
    \begin{table}
    \begin{center}
    \begin{tabular}{c|ccc}
       Obs. wavelength     & $\Delta \alpha'=100\,$mas & $200\,$mas & $400\,$mas \\
    \cmidrule{1-4}
    $0.6263\,\mu$m & 8.1\,mag & 11.0\,mag & 11.2\,mag \\
    $1.245\,\mu$m  & 7.0\,mag & 12.3\,mag & 13.7\,mag \\
    $2.182\,\mu$m  & 6.6\,mag & 7.6\,mag & 11.2\,mag \\
    \end{tabular}
    \caption[xxx]{SPHERE detection limits derived with the SPHERE ETC for three observing wavelengths as functions of the angular separation $\Delta \alpha'$ to the reference star. For details, see Sect.~\ref{sec:sphere}.}
    \label{tab:detection_limits}
    \end{center}
    \end{table}

\begin{figure}
        \centering
        \includegraphics[width=\hsize]{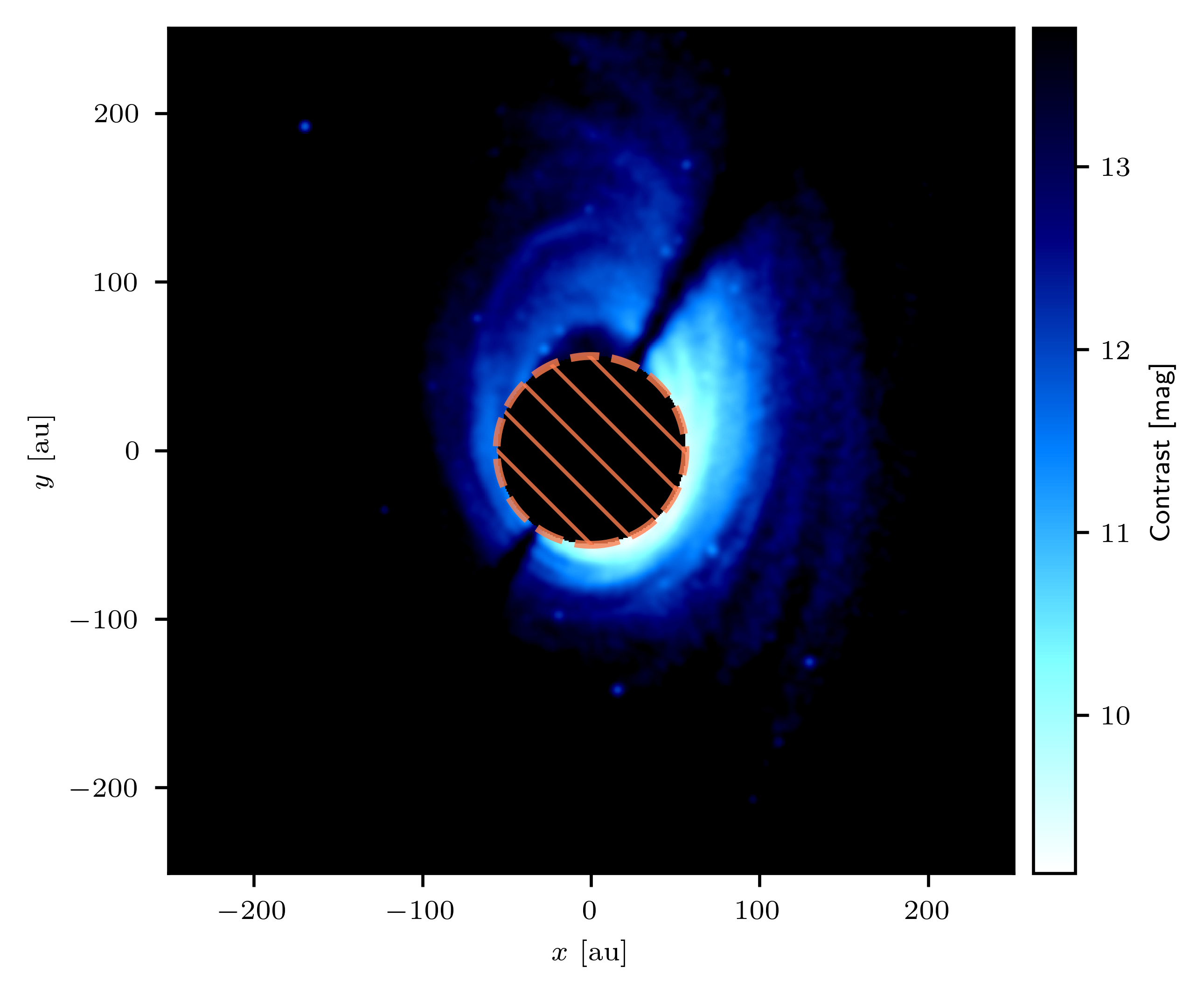} 
            \caption{Contrast map corresponding to a synthetic observation at $ 1.245\,\mu$m based on run 5 at 150 kyr assuming a distance of $140\,$pc. For illustrative purposes, the hatched area of radius $400\,$mas in the center is masked, and the color bar is limited to the maximum range of detectable contrast values outside the masked region.}
            \label{fig:example_sphere_observation}
    \end{figure}

    % results
    On the basis of these detection limits, we find that in many cases, the shadows induced by the misalignment of the inner and outer disk can be imaged directly using SPHERE. Figure \ref{fig:example_sphere_observation} exemplarily shows a contrast map based on a synthetic observation of a system at a distance of $140\,$pc, which is comparable to the distance to SU Aur \citep{2018A&A...616A...1G}, observed at $1.245\,\mu$m. Moreover, the central 400\,mas in the image are masked, and the displayed color bar is clipped at 13.7\,mag, which is the lowest detection limit outside the masked region, see Table \ref{tab:detection_limits}. As a result, this depiction of the observation is rather conservative, meaning, slightly fainter features might be visible in an equivalent real observation. The image exhibits clear features of the expected shadows extending in almost diametrically opposed directions approximately from the center of the image. Moreover, even when increasing the distance of the simulated system to $400\,$pc, these features remain. Overall, we find the obtained contrast values of such shadow features to be in the detectable range for systems with various relative orientations between the inner and outer disk observed at different inclination angles. Comparing observations performed using different filters, see left column of Fig.~\ref{fig:alma_multi_bands} in the Appendix, we furthermore find that these features seem to be particularly pronounced in the NIR wavelength range and less so in the VIS wavelength range. For the latter case, the star and the inner disk strongly dominate the total flux of the system, such that contrast values in regions of the outer disk are often outside the detectable range. To enable the comparison of observations at different wavelength, the displayed synthetic observations in the VIS/NIR and submm/mm wavelength ranges in this figure were all generated assuming the same distance of $d=140\,$pc. Interestingly, the synthetic observations with IRDIS at $1.245\,\mu$m often lead to more pronounced features than comparable observations at $2.182\,\mu$m, which can be attributed to the better contrast detection limits provided by SPHERE in the J band. Therefore, we conclude that scattered light observations, particularly in the NIR, allow us to directly image these shadows when using SPHERE. Next, we analyze the feasibility of observing shadows in the submm/mm wavelength range using ALMA.

\subsection{Dust cooling and heating}
\label{sec:cooling_and_heating}

As shown in the ideal flux maps in Fig.~\ref{fig:alma_m154_idealVScasa}, direct stellar radiation cannot illuminate the entire outer disk evenly because of the shadowing effects caused by the inner disk. Consequently, these shadowed regions are heated indirectly only by scattered light and re-emitted radiation from the illuminated regions of the outer disk, leading to a reduced dust temperature compared to the primarily directly heated regions of the outer disk.\footnote{Figure \ref{fig:temp_dist} in the appendix shows exemplarily the temperature distribution across two vertical cuts through the mid-plane of the outer disk, one corresponding to a shadowed region and the other to a directly heated region.}

In the following, we aim at investigating whether the resulting discontinuity in the spatial temperature distribution can be detected with ALMA. However, our synthetic RT observations represent merely a snapshot and cannot account for the dynamical processes of hot dust material entering the shadowed regions.  Thus, before validating the feasibility of such a shadow detection, we have to evaluate the potential impact of dynamic processes on the dust temperature in the shadowed regions. If for instance the dust cooling is inefficient, the hot material may not reach a lower temperature within the transit time of the shadow for the effect to become detectable. The defining factor is the cooling time of dust grains itself. 

A similar model was explored in \cite{2019MNRAS.486L..58C}. However, in their model a dust-gas interaction was assumed in order to determine the temporal evolution of the dust temperature within the shadowed region. In contrast to \cite{2019MNRAS.486L..58C}, we assume that the dust is sufficiently far away from the central star. Hence, the heating and cooling processes are dominated by radiation and not by gas-dust interactions such as viscous heating. Such conditions can be expected to prevail in the outer disk. As a result, the temporal evolution of the dust temperature $T_{\mathrm{d}}$ simply follows
\begin{equation}
    \frac{\mathrm{d} T_{\mathrm{d}}  }{\mathrm{d} t} = \frac{\overline{\kappa}_{\mathrm{abs}}(T_{\mathrm{d}})}{C_{\mathrm{v}}(T_{\mathrm{d}})} \left(  T_{\mathrm{RT}}^4-T_{\mathrm{d}}^4   \right)
    \label{TEvoulution}
\end{equation}
where 
\begin{equation}
    \overline{\kappa}_{\mathrm{abs}}(T_{\mathrm{d}})=\frac{\pi}{\sigma T_{\mathrm{d}}^4} \int{ \kappa_{\mathrm{abs,\lambda}} B_{\lambda}(T_{\mathrm{d}}) \mathrm{d} \lambda      }
    \label{PlanckMean}
\end{equation}
is the Planck mean opacity, $\sigma$ the Stefan–Boltzmann constant, and $T_{\mathrm{d}}$ the dust temperature in the non-shadowed outer disk calculated with the MC approach. The opacity of absorption $\kappa_{\mathrm{abs,\lambda}}$ corresponds to that of the optical properties of the dust model outlined in Sect.~\ref{sec:radiative_transfer_pos_processing}. For the heat capacity $C_{\mathrm{v}}(T_{\mathrm{d}})$ of the grain material, we use the spline interpolated values of the data presented in \cite{DraineLi2001}.

\begin{figure}
    \centering
    \includegraphics[width=\hsize]{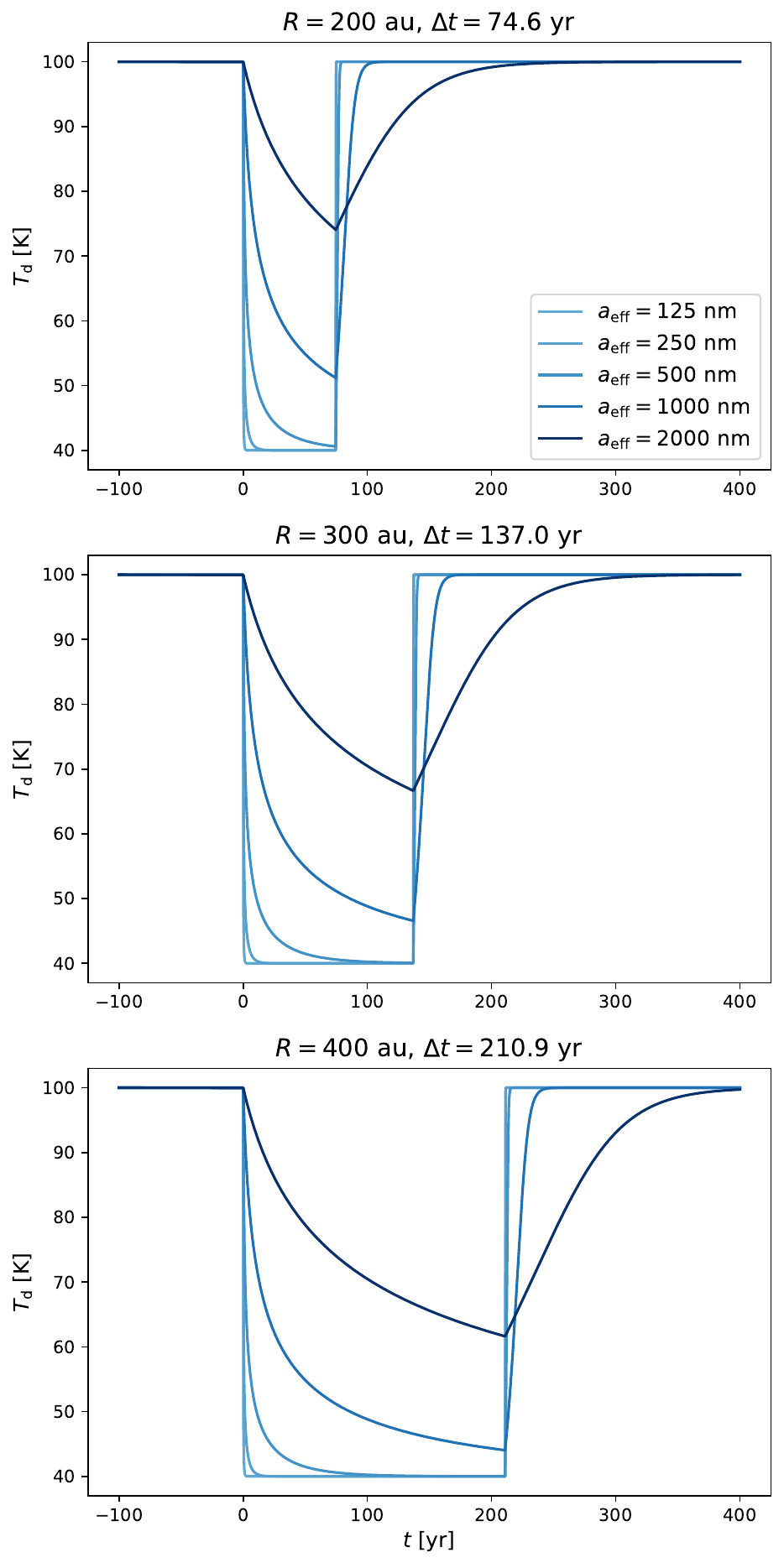} 
        \caption{Temporal evolution of the dust temperature $T_{\mathrm{d}}$ within the outer disk for different grain sizes $a_{\mathrm{eff}}$ and shadow transition periods $\Delta t$. Calculating are performed for the three distances from the central star of $R=200\, \mathrm{au}$ (top panel), $R=300\, \mathrm{au}$ (middle panel), and $R=400\, \mathrm{au}$ (bottom panel), respectively.  }
        \label{fig:RTheating}
\end{figure}

In Fig.~\ref{fig:RTheating} we show the resulting dust cooling process for three exemplary distances $R$ from the central star assuming a Keplerian rotation of the outer disk and a typical opening angle of $15^\circ$ for the shadowed region (see upper plot in Fig.~\ref{fig:SU_Aur}). At that distance, the MCRT calculations predict a temperature decline from $\approx 100\ \mathrm{K}$ in the fully illuminated outer disk down to $\approx 40\ \mathrm{K}$ in the shadowed regions. Under such conditions, grains with sizes $a_{\mathrm{eff}} \leq 125\ \mathrm{nm}$ can efficiently cool down to the lowest temperatures before reaching the end of the transit period. Only the largest but least abundant grains with sizes of $a_{\mathrm{eff}} \geq 500\ \mathrm{nm}$ cannot fully cool down to the lowest temperature. For the following analysis, we therefore make the assumption that the transit time does suffice for efficient cooling to take place inside the shadowed regions. Hence, we approximate the underlying temperature distributions of our models with the equilibrium temperature distributions calculated using MCRT simulations.

\subsection{ALMA observations}
\label{sec:alma}
The ideal flux maps of all considered configurations were processed with the tool Common Astronomy Software Applications (CASA), using version 6.4.1 \citep{2007ASPC..376..127M}. We generated synthetic ALMA observations for $\lambda=850\,\mu$m (band 7) and $1300\,\mu$m (band 6) assuming the celestial coordinates of the reference star as well as a distance of $140\,$pc. These synthetic observations were performed for configurations C43-4, C43-5, C43-6, and C43-7 using the \texttt{simobserve} task of CASA, assuming a rather long total observation time of $\Delta t_{\rm obs}=12$\,h in order to obtain high quality results. We note, that such a long observation time can not be acquired with ALMA during a single night for the considered reference star, however, as will be shown in the following, the feasibility of a shadow detection is limited by the sensitivity which can be achieved by performing multiple observations. The combination of these bands and configurations results in angular resolutions ranging from about $\theta_{\rm res}=0.06-0.4\,$arcsec \footnote{See ALMA technical handbook, Table 7.1: \\\url{https://almascience.eso.org/documents-and-tools/cycle9/alma-technical-handbook}}, which is comparable to the width of the shadow features found in our ideal flux maps at different radial distances from the star. Subsequently, the \texttt{tclean} task of CASA was performed to reconstruct a sky model, using the spectral mode \texttt{mfs} and \texttt{natural weighting}, in order to achieve the maximum imaging sensitivity. The threshold was set to the values that were determined with the ALMA Sensitivity Calculator \footnote{\url{https://almascience.eso.org/proposing/sensitivity-calculator}}. We find that due to the high level of noise, it is generally difficult to infer the presence of a shadow simply by visually inspecting these synthetic observations. While the sensitivity of ALMA does suffice to detect the inner disk in all of the considered cases, only a fraction of the simulated maps show features that originate from further extended structures. The features may include an outer disk that is separated from an inner disk through a gap, arc-like structures that are often close to the inner rim of the outer disk, spiral arms, and rarely even a low contrast shadow in the region of the outer disk. We note in passing, that supplementary tests have been performed, where the considered maximum grain size of dust grains in the inner disk has been artificially increased to $1\,$mm. Notably, this change did not appear to have a significant qualitative impact on the simulation outcomes. As a result, it will not be taken into consideration in the subsequent analysis.

\begin{figure}
   \centering
   \includegraphics[width=\hsize]{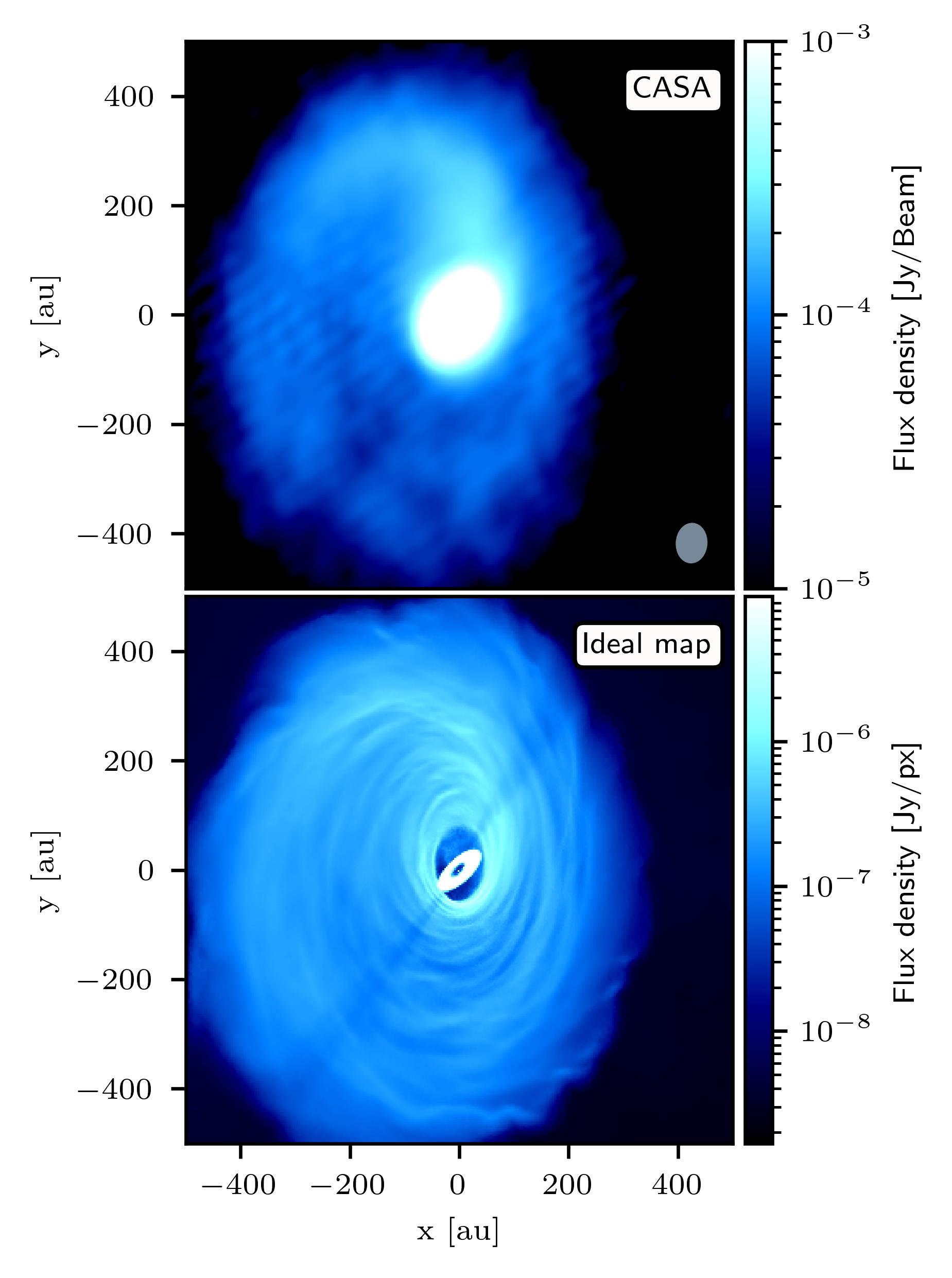}
      \caption{Top: Synthetic ALMA observation at $850\,\mu$m using configuration C43-4 (rms=$3.32\cdot 10^{-4}\,$Jy/Beam) based on run 5 at 150 kyr assuming a distance of $140\,$pc. The beam (gray) is displayed in the lower right corner of the plot. Bottom: Corresponding ideal flux map. 
              }
         \label{fig:alma_m154_idealVScasa}
   \end{figure}

An analysis of our results suggests, that for the shadows to be detectable the system has to be misaligned and formed with an infall angle of $\alpha\neq 0$. Otherwise, a detection with ALMA seems unfeasible for all investigated systems. Moreover, we find that the best combination of considered observing wavelength and configuration depends primarily on the size of the reconstructed beam, i.e., it requires a trade-off between sensitivity and resolution. In our case, this corresponds to a combination of the shorter wavelength with a compact ALMA configuration or the longer wavelength with a wider configuration. Figure \ref{fig:alma_m154_idealVScasa} exemplarily shows a reconstructed synthetic ALMA observation using configuration C43-4 that exhibits, alongside an extended streamer feature, such a shadow feature (top plot) as well as the corresponding ideal flux map (bottom plot). We note that the detection of this low contrast feature required a careful inspection of the map, as well as clipping of the color bar due to the high brightness of the inner disk. The width of the shadow feature appears to be smeared out to approximately the size of the beam. In the case of this system, it could only be found considering this rather compact configuration, while the more extended configurations C43-5 to C43-7 lead to non-detections, which will be further discussed in the context of instrument sensitivity in Sect.~\ref{sec:alma_sensitivity}. In another test, we confirmed that even the very long baseline configurations C43-8 and C43-9, which could potentially be useful for detecting shadows by resolving the inner regions of the outer disk where the disk is bright, did not result in a successful detection either. Finally, we find that the more evolved systems (e.g., $150\,$kyr) more often resulted in detectable shadow features than younger systems ($50\,$kyr), which is exemplarily shown in Fig. \ref{fig:alma_m138_146_154} in the appendix. 

\subsubsection{Instrument sensitivity}
\label{sec:alma_sensitivity}
The fact that most of these shadow features which can clearly be seen in ideal flux maps are not visible in corresponding synthetic simulations can be explained by a too small ratio of the flux emitted by the outer disk to the sensitivity provided by ALMA. To verify this, we assessed the feasibility of detecting the shadows for the system shown in Fig.~\ref{fig:alma_m154_idealVScasa} with future instrumentation that offers an improved sensitivity. This was done by artificially increasing the values of the corresponding ideal flux map pixelwise by a factor of 10 before post-processing it with CASA. The results of this analysis for configurations C43-5 (upper row), C43-6 (middle row), and C43-7 (bottom row) can be found in Fig.~\ref{fig:alma_m154_CASA_vs_CASAx10} (Appendix), where the left column depicts the results corresponding to the original unaltered ideal flux map and the right column the results after the artificial increase of flux values. We find that this increase in flux, which is equivalent to a likewise improvement of instrument sensitivity, is already sufficient for the emergence of shadow features in observations using any of the four considered configurations. Additionally, it is worth mentioning that at a distance of 140\,pc, only configuration C43-4 provides a maximum recoverable scale (MRS) for both wavelengths large enough to encompass the inner disk and a part of the outer disk, which is not the case for any of the other considered much wider configurations. However, since the change in orientation of the disks and consequently the position of the shadow is negligible on timescales of a few years or decades \citep[figures 12 and 13 in][show that the change in orientation is less than $1^{\circ}$ on timescales of a few thousand years]{Kuffmeier2021}, it is possible to combine observations in the more extended configurations with observations in the compact configurations to obtain a better UV coverage. Figure \ref{fig:combination_alma_C4plus7} in the appendix exemplarily shows the results of a multi-configuration observation, combining interferometric data of a high-resolution observation using configurations C43-7 ($1 \times \Delta t_{\rm obs}$) with data obtained for configuration C43-4 ($0.23 \times \Delta t_{\rm obs}$), which covers a sufficiently large angular scale to encompass the whole system. As a result, the addition of configuration C43-4 data improves the synthetic observation, allowing for a visual detection of a weak shadow feature in the lower left quadrant of the image.

\subsubsection{More distant systems}
\label{sec:alma_larger_dist}
\begin{figure}
   \centering
   \includegraphics[width=\hsize]{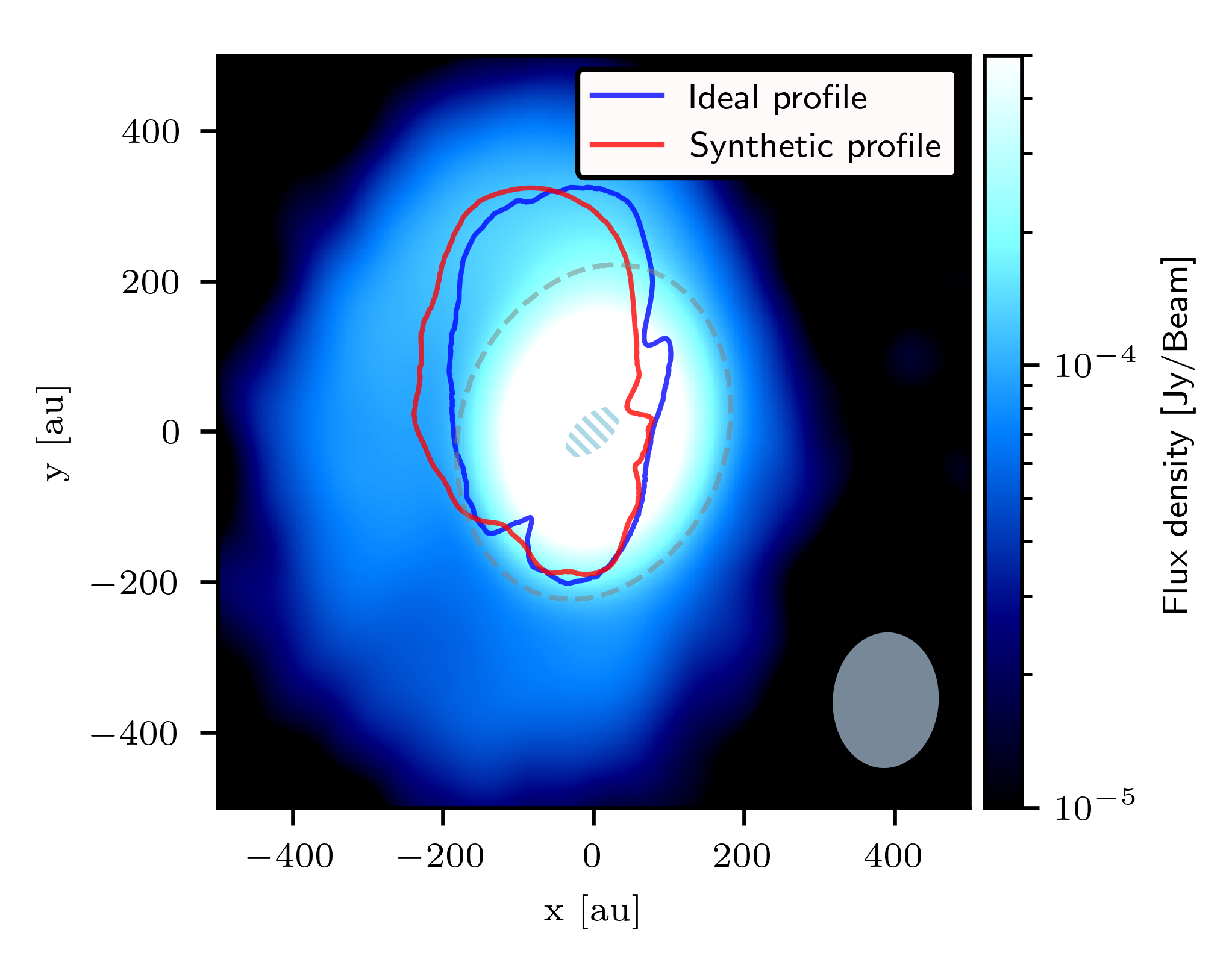}
      \caption{Synthetic ALMA observation at $850\,\mu$m using configuration C43-4 (rms=$1.59\cdot 10^{-4}\,$Jy/Beam) based on the simulation of run 5 at 150 kyr assuming a distance of $400\,$pc. Azimuthal profiles for the ideal flux map (red) and the synthetic observation (blue) are shown as solid lines. The hatched central region indicates the approximate region of the inner disk as depicted in the corresponding ideal flux map, which is furthermore enclosed by a gray dashed line, which marks the area within one beam size (FWHM) from the approximate inner disk. The beam (solid gray area) is displayed in the lower right corner of the plot. For details, see Sect.~\ref{sec:alma_az_profiles}.}
         \label{fig:alma_m154_400pc_wAzimuthalProfiles}
   \end{figure}

We investigated the feasibility of detecting these shadows for more distant systems using configuration C43-4. Figure \ref{fig:alma_m154_400pc_wAzimuthalProfiles} shows such a synthetic observation of the same system as investigated before (see Fig.~\ref{fig:alma_m154_idealVScasa}), but now at an increased distance of $400\,$pc. At this distance, the shadow feature is significantly narrower in the ideal flux map and, thus, appears notably smeared out in the synthetic observation, leading to a much smaller observed flux reduction in the region of the shadow. As a result, it can barely be seen in Fig.~\ref{fig:alma_m154_400pc_wAzimuthalProfiles}, even after clipping the color bar. This is especially the case for the top right quadrant, where it seems to have completely vanished. Hence, we conclude that although the likelihood of detecting shadows in systems at greater distances appears reduced, the potential for the spatial extension of these shadows, possibly spanning hundreds or even thousands of astronomical units, still allows for feasible detections. 

\subsubsection{Detection tool: Azimuthal profiles}
\label{sec:alma_az_profiles}
Due to the high level of noise present in the reconstructed ALMA observations (see for instance Fig.~\ref{fig:alma_m154_400pc_wAzimuthalProfiles}) and the generally low flux level in the shadowed region, it is often difficult to infer the presence of the shadow solely based on a visual inspection of the observation. However, while the flux value in the synthetic observation may, due to underlying noise, strongly deviate from its value in the ideal flux map, the process of integrating over a region in the map can effectively mitigate this issue, resulting in a smaller relative deviation. Since the shadow is cast onto the outer disk, extending in approximately radial direction from the inner edge of the disk to its far-out regions, it leaves a characteristic feature in the form of two almost diametrically opposed dips in the azimuthal brightness profile. Here, an azimuthal profile is computed by radially integrating the flux map along different azimuthal directions, $\phi\in\left[0,\,2\pi \right)$, in the plane of the sky, starting from the center of the map. 

In order to calculate the azimuthal profile, we use CartToPolarDetector\footnote{\url{https://github.com/anton-krieger/CartToPolarDetector}} \citep{2022A&A...662A..99K}, a tool that precisely converts a Cartesian detector to its polar representation. For the obtained polar flux maps, we use $360\times720$ polar pixels (in azimuthal $\times$ radial directions) and a detector radius that corresponds to the specific MRS of the underlying configuration and wavelength of the map. Since the inner disk is significantly brighter than the outer disk, a central region of the flux map is masked. In order to avoid integrating noise-dominated regions in the map, it is furthermore necessary to limit the radial extent ($\Delta r$) of the integrated region. 

Figure \ref{fig:alma_m154_400pc_wAzimuthalProfiles} shows the azimuthal profiles obtained from the corresponding ideal flux map (blue) and the synthetic observation (red). For the latter, we limited the integrated region to a radial range of $\Delta r=300\,$au. Furthermore, we assumed an inner disk size and shape (hatched central region) that approximates the inner disk as seen in the ideal flux map, and masked the inner region (inside the gray dashed line) according to one FWHM of the synthetic ALMA beam. A detailed description of our method to calculate these azimuthal profiles is presented in Sect.~\ref{sec:app:azimuthal_profiles} of the Appendix. 

For every direction $\phi$, the distance of the displayed azimuthal profile to the center of the map is proportional to the functional value of the azimuthal profile in that specific direction. We note, that the depiction of these profiles in polar coordinates in combination with the synthetic observation in the background is chosen as it allows for a simple visual recognition of their common patterns. Nonetheless, we also included a plot of these azimuthal profiles as a function of the azimuthal angle in Fig.~\ref{fig:traditional_azimuthal_profiles} (Appendix). The ideal azimuthal profile in Fig.~\ref{fig:alma_m154_400pc_wAzimuthalProfiles} shows two clear dips that are almost diametrically opposed, which match the direction of the shadow features found in the ideal flux map. The determined synthetic profile exhibits a very wide and shallow dip in the lower left quadrant, and an apparently slightly displaced dip in the upper right quadrant. This result suggests, that the information of the existence of a shadow in this quadrant may potentially be retrieved by the usage of our method. However, additional dips can be found in the azimuthal profile that do not match other corresponding shadow features, which complicates the interpretation of these dips. Generally, these dips can be attributed to the overall low level of flux of the outer disk compared to the provided instrument sensitivity. However, even though the interpretation may be ambiguous, we conclude that the determination of azimuthal profiles is a useful tool that can be used to support claims regarding detected shadow features. In particular, it can help find very shallow shadow features that may otherwise be overlooked when visually inspecting the observations, as demonstrated by the results in Fig.~\ref{fig:alma_m154_400pc_wAzimuthalProfiles}. 

\section{Summary and conclusions}
    % Goal: we set out to investigate feasibility -> Method to generate synthetic observations
    We investigated the feasibility of detecting shadows in observations of circumstellar disks around young stellar objects, that form as a consequence of a late infall event. It is based on snapshots of previously performed hydrodynamical simulations of a star hosting a circumstellar disk that collides with a cloudlet \citep[][]{Kuffmeier2021}. These simulations were subsequently post-processed using MC radiative transfer simulations to calculate wavelength-dependent ideal flux maps in the NIR/VIS and the submm/mm wavelength range (see Sect.~\ref{sec:radiative_transfer_pos_processing}), which exhibit clear shadow features that extend across the outer disk. In order to assess the feasibility of detecting such shadows with real observations, synthetic observations were generated from these maps, assuming the usage of the instruments SPHERE/VLT and ALMA for the VIS/NIR and submm/mm wavelength range, respectively. In short, this involved convolving the ideal maps with a beam of wavelength-dependent size in the case of SPHERE observations (see Sect.~\ref{sec:sphere}) as well as the simulation of an observation and subsequent image reconstruction using CASA in the case of ALMA (see Sect.~\ref{sec:alma}). The thereby obtained synthetic observations served as a starting point from which the following conclusions were drawn. Figure \ref{fig:alma_multi_bands} in the Appendix shows synthetic observations exemplarily for all considered observing wavelengths of SPHERE and ALMA assuming a distance of $d=140\, \mathrm{pc}$ to allow for a direct comparison of observations at different wavelengths. We find, that shadow features can be observed both via scattered light observations with SPHERE and via thermal emission maps with ALMA. 
    % findings SPHERE
    First, we recall the key points of our analysis of our derived synthetic SPHERE observations.
    \begin{itemize}
        \item We applied realistic contrast detection limits obtained from the ETC and assumed a distance of 140\,pc to the simulated system. Corresponding synthetic observations exhibit various features such as gaps, arcs, streamers, and pronounced shadows.
        \item We find the contrast between the shadowed region in the outer disk and neighboring regions to be well inside the detectable range for systems with various relative orientations between the inner and outer disk observed at different inclination angles.
        \item Observations in the NIR wavelength range are generally well suited for a detection of shadows. Our analysis suggests that observations in the J band with the instrument IRDIS at $1.245\,\mu$m can be expected to yield the most promising results. On the contrary, a detection in the VIS wavelength range seems to be hindered due to a relatively high stellar flux.
    \end{itemize}
    % findings ALMA
    Next, we recall the key points of our analysis of the obtained synthetic ALMA observations.
    \begin{itemize}
        \item We estimate, that for grains of sizes $a_{\mathrm{eff}} \leq 250\ \mathrm{nm}$, which likely represent the majority of all available grains in the shadowed region of the outer disk, the shadowing time does suffice for efficient cooling, see Sect.~\ref{sec:cooling_and_heating}. 
        \item Maps were reconstructed for simulated observations in two bands (6 and 7) and four configurations (C43-4 to C43-7), assuming a distance of 140\,pc to the systems. 
        \item Reconstructed maps exhibited various features, including gaps, arcs, streamers, and rarely low contrast shadows. 
        \item Our results suggest, that detecting a shadow is a challenging task as the flux of the outer disk is rather low compared to the provided sensitivity of the instrument, which leads to noisy reconstructed maps. The most promising results were obtained using a combination of band 7 (at $850\,\mu$m) and the rather compact configuration C43-4. In the rare cases, which allow for a shadow detection, careful inspection of the map is mandatory due to the high brightness of the inner disk. 
        \item We find that for the impact of the shadows to be detectable, the system has to be misaligned and formed with an infall angle of $\alpha\neq 0$. Furthermore, if the system has not yet evolved for a sufficient amount of time ($\leq50\,$kyr), a detection is rather unlikely.
        %\item Furthermore, an improvement of the provided instrument sensitivity by one order of magnitude would suffice to allow for shadow detections of various of the considered systems for multiple configurations, meaning, future instrumentation may allow us to detect a much greater abundance of shadows than have been reported thus far (see Sect.~\ref{sec:alma_sensitivity}). 
        \item Detecting shadows in systems at a (greater) distance of 400\,pc, while more challenging, may still be possible for certain systems (see Sect.~\ref{sec:alma_larger_dist}). We therefore investigated the possibility of retrieving the information of the presence of a shadow with these reconstructed maps based on azimuthal profiles. We find that these profiles can aid the search for very shallow shadow features that may be overlooked by just visually inspecting the observations. 
    \end{itemize}
    % general remark
    We conclude that an event of late infall of material onto the system of a star hosting a disk has the potential to cause shadow features that are observable with currently available instruments. In particular, this is the case with high-contrast and high-sensitivity instruments such as SPHERE and ALMA in the VIS/NIR and submm/mm wavelength range, respectively. Hence, the scenario of late infall does offer a plausible explanation for the emergence of shadows that are observed in real systems.

\section*{ORCID iDs}
A. Krieger \orcidlink{0000-0002-3639-2435}
\href{https://orcid.org/0000-0002-3639-2435}
     {https://orcid.org/0000-0002-3639-2435}\\
M. Kuffmeier \orcidlink{0000-0002-6338-3577}
\href{https://orcid.org/0000-0002-6338-3577}
     {https://orcid.org/0000-0002-6338-3577}\\
S. Reissl \orcidlink{0000-0001-5222-9139}
\href{https://orcid.org/0000-0001-5222-9139}
     {https://orcid.org/0000-0001-5222-9139}\\
C. P. Dullemond \orcidlink{0000-0002-7078-5910}
\href{https://orcid.org/0000-0002-7078-5910}
     {https://orcid.org/0000-0002-7078-5910}\\
C. Ginski \orcidlink{0000-0002-4438-1971}
\href{https://orcid.org/0000-0002-4438-1971}
     {https://orcid.org/0000-0002-4438-1971}\\
S. Wolf \orcidlink{0000-0001-7841-3452}
\href{https://orcid.org/0000-0001-7841-3452}
     {https://orcid.org/0000-0001-7841-3452}

\begin{acknowledgements}
A.K. and S.W. acknowledge the support of the DFG priority program SPP 1992 "Exploring the Diversity of Extrasolar Planets (WO 857/17-2)". This research was supported in part through high-performance computing resources available at the Kiel University Computing Centre.
The research of M.K. is supported by H2020 Marie Sklodowska-Curie Actions (897524) and a Carlsberg Reintegration Fellowship (CF22-1014). S.R. acknowledges funding from the European Research Council via the ERC Synergy Grant ``ECOGAL'' (project ID 855130), from the German Excellence Strategy via the Heidelberg Cluster of Excellence (EXC 2181 - 390900948) ``STRUCTURES''. S.R. also thanks for computing resources provided by {\em The L\"{a}nd} and DFG through grant INST 35/1134-1 FUGG and for data storage at SDS@hd through grant INST 35/1314-1 FUGG. 
\end{acknowledgements}

\bibliography{general}
\bibliographystyle{aa}

%\appendix

\begin{appendix}

\section{Azimuthal profile determination}
\label{sec:app:azimuthal_profiles}
%%%%%%%%%%%%%%%%%%%%%%%%%%%%%%%%%%%%%%%%%%%%%%%%%%%%%%%%%%%%%%%%%%
   \begin{figure}[!htb]
   \centering
   \includegraphics[width=\hsize]{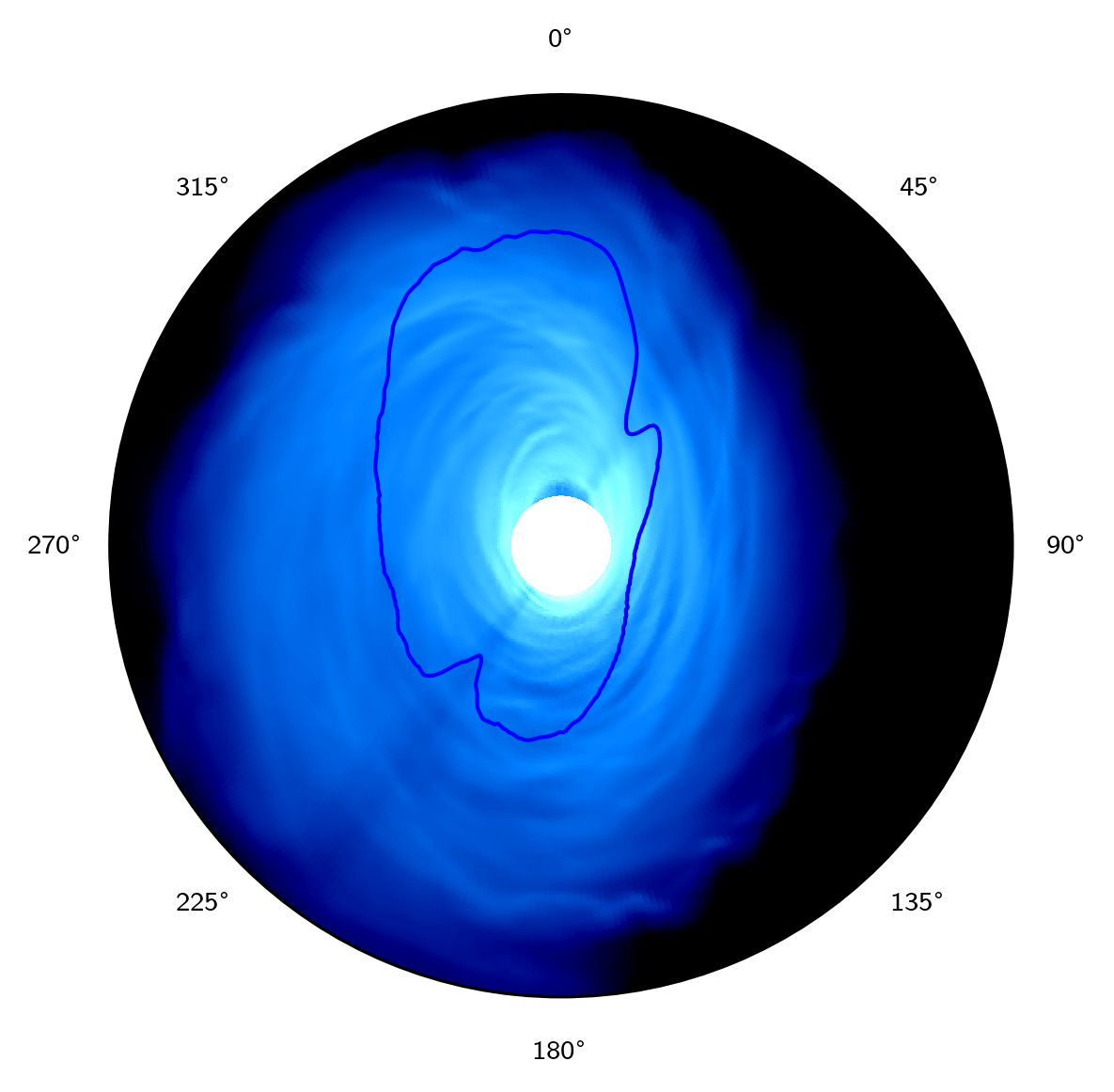}
      \caption{% previously: m, w, c = 42, 850, 4, now it changed
              Ideal flux map at $850\,\mu$m based on run 5 at $150 \,$kyr. The blue curve shows the corresponding azimuthal profile, which was obtained after masking the central circular white region. 
              }
         \label{fig:ideal_map_with_azimuthal_profile}
   \end{figure}
   
In the following, we present a method to determine and to analyze azimuthal profiles in order to detect radially extending shadow features in observations. All analyses presented in this section are based on run 5 at $150 \,$kyr assuming a distance of $400\,$pc. Figure \ref{fig:ideal_map_with_azimuthal_profile} shows the corresponding ideal flux map for this system at $850\,\mu$m, overlayed with its ideal azimuthal profile (blue curve).

 The shape of the masked central region (white region) is circular and chosen such that it encompasses the whole inner disk. For every direction $\phi$, the displayed distance of the azimuthal profile to the origin of the polar grid is proportional to the functional value of the azimuthal profile in that specific direction. For all our models, we find that the locations of both shadow features in the flux map clearly match the positions of dips in the ideal azimuthal profile.

\FloatBarrier
\subsection{Masking methods}
In the case of the synthetic ALMA observations, on the contrary, a more sophisticated masking method has to be used to calculate an azimuthal profile that exhibits dips that match both locations of the shadow without introducing additional features that may lead to a false or no shadow detection. In general, there are several factors that make the calculation of an azimuthal profile which serves this purpose difficult: the low flux level of the outer disk, a small flux difference between the regions inside and the regions just outside the shadow, the shape of the synthesized beam, the inclination of both the inner and the outer disk, and the fact that the FWHM of the beam and the inner disk are of similar angular extent.

Therefore, we tested different post-processing methods for retrieving the information of the presence of shadows from the synthetic ALMA observations. In the end, however, none of them were able to unambiguously lead to a detection, which is for the most part a consequence of the generally low flux level of the outer disk. Nonetheless, we were able to make a comparison and identified the method that resulted in the best azimuthal profiles for detecting shadows in disks and can be used as a tool to find shallow shadow features. 

Before describing the method in Sect.~\ref{sec:app:shadow_retrieval}, it is worth mentioning the various other approaches that generally lead to unwanted features in the azimuthal profile, making its appearance differ strongly from their ideal counterpart. An inapt approach would often either lead to a lack of dip features in the azimuthal profile or to a high abundance of them with no clear pattern. For instance, we find that it does not suffice to mask a circular region, as this is neither taking into account the specific shape and size of the inner disk nor of the beam. When instead masking a region that has the same shape as the synthesized beam, i.e., of its main lobe, and testing different total angular extents of the masked region, the azimuthal profile still exhibits unwanted features. This is most likely due to the fact, that in our models the inner disk has a size that is comparable to that of the beam, which leads to a spreading of its flux that does not match the shape of the beam alone. In another method that we tested, a certain number of the brightest pixels in the centers of the flux maps were removed. However, even when restricting the successive removal of the brightest pixels to those that are neighbors of previously removed bright pixels, this method leads to masked regions with very complex structures and various features in the azimuthal profile. This is partly due to the noise and partly due to the fact that the flux of the inner disk is spread according to the PSF such that it overlaps with the flux of the inner edge of the outer disk. Due to the former effect, there are undesired additional features in the azimuthal profile, and because of the latter, it can be expected that the dips, which indeed are a consequence of the shadows, become shallower and wider, thus decreasing the likelihood for a shadow detection. 

% inner disk is almost an ellipse, lets use this information to derive a better procedre
\FloatBarrier
\subsection{Shadow retrieval}
\label{sec:app:shadow_retrieval}
In order to generate the most suitable azimuthal profiles for detecting shadows in disks, we made use of the fact, that the shape of the inner disk strongly resembles an ellipse, as can exemplarily be seen in Fig.~\ref{fig:example_sphere_observation}. The method can then be summarized as follows: Firstly, we make guesses regarding the shape and size of the inner disk by approximating it by an ellipse, and calculate a beam-size and beam-shape dependent masking region that encompasses the guessed inner disk. Secondly, for different areas of the masked regions, we determine the best guessed disk sizes and shapes. Thirdly, for these cases we compute normalized azimuthal profiles, that result in the most reliable shadow predictions. 

\begin{figure}
   \centering
   \includegraphics[width=\hsize]{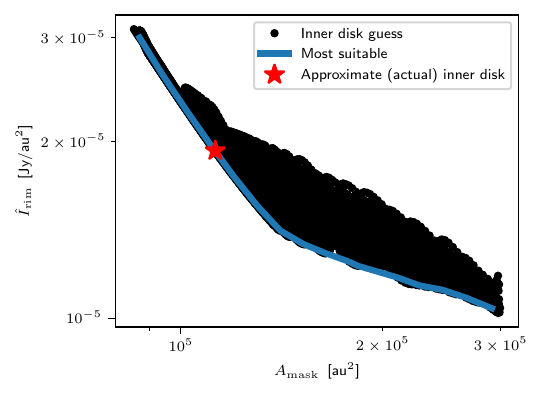}
      \caption{Distribution of obtained data points corresponding to different guessed inner disk shapes and sizes (black dots) in the $A_{\rm mask}$-$\hat{I}_{\rm rim}$ plane based on a synthetic ALMA observation of run 5 at $150 \,$kyr and $850\,\mu$m. The blue curve marks the position of most suitable inner disk guesses. A red star highlights the position corresponding to an inner disk of approximately the shape and size of the (actual) inner disk as seen in the ideal flux map. 
              }
         \label{fig:Amask_Ihatrim}
   \end{figure}

In particular, we test different guesses for the inner disk by successively increasing its area, varying its eccentricity, and changing the orientation of its major axis. For any given inner disk guess, the masked area is calculated by extending the region of the guessed inner disk by at least the FWHM of the main lobe of the beam, i.e., we calculate the area that is covered by the beam, when moving it along the outer edge of the guessed inner disk. By doing so, the masked area is adjusted according to the shape and size of both the inner disk and the beam. We note, that if the noise level is not too high, it can be beneficial to use 2 to 4 times the FWHM instead of just one FWHM, as this further reduces the impact of the inner disk on the determined azimuthal profiles. For each tested guessed inner disk, we then determine the area of the masked region $A_{\rm mask}$, the flux of a one polar pixel wide rim region that surrounds the masked region $I_{\rm rim}$, as well as the total area of this rim region $A_{\rm rim}$. As a result, each tested guessed inner disk corresponds to a point in the $A_{\rm mask}$-$\hat{I}_{\rm rim}$ plane, with $\hat{I}_{\rm rim}=I_{\rm rim}/A_{\rm rim}$, as can be seen in Fig.~\ref{fig:Amask_Ihatrim}, where each black dot represent a different inner disk guess. Next, we define the most suitable inner disk guesses as those, which minimize $\hat{I}_{\rm rim}$ for any given value of $A_{\rm mask}$ (blue curve in Fig.~\ref{fig:Amask_Ihatrim}). 

It is worth mentioning, that maximizing the flux of the masked area for a fixed guessed inner disk area does not work for our purpose, since this approach tends to favor unrealistically high eccentricities close to unity for the guessed inner disks. This behavior can be explained by the fact, that the masked area strongly increases with the eccentricity of the disk, such that these highly eccentric disks result in extremely stretched masked regions with very high $A_{\rm mask}$ values.

% added more points to the by scaling the beam only and assuming no inner disk
Furthermore, we determined the approximate parameters of an ellipse that resembles the shape and size of the actual inner disk based on its ideal flux map (see bottom plot in Fig.~\ref{fig:alma_m154_idealVScasa}) and computed the position of the corresponding point in the $A_{\rm mask}$-$\hat{I}_{\rm rim}$ plane, which is shown as a red cross in Fig.~\ref{fig:Amask_Ihatrim}. Based on that we verified, that the approximate ideal disk shape and size indeed corresponds to a point in the $A_{\rm mask}$-$\hat{I}_{\rm rim}$ plane that lies on the curve of most suitable inner disk guesses, for all considered models, configurations, and wavelengths. 

\begin{figure*}
   \resizebox{\hsize}{!}
            {\includegraphics[width=\textwidth,clip]{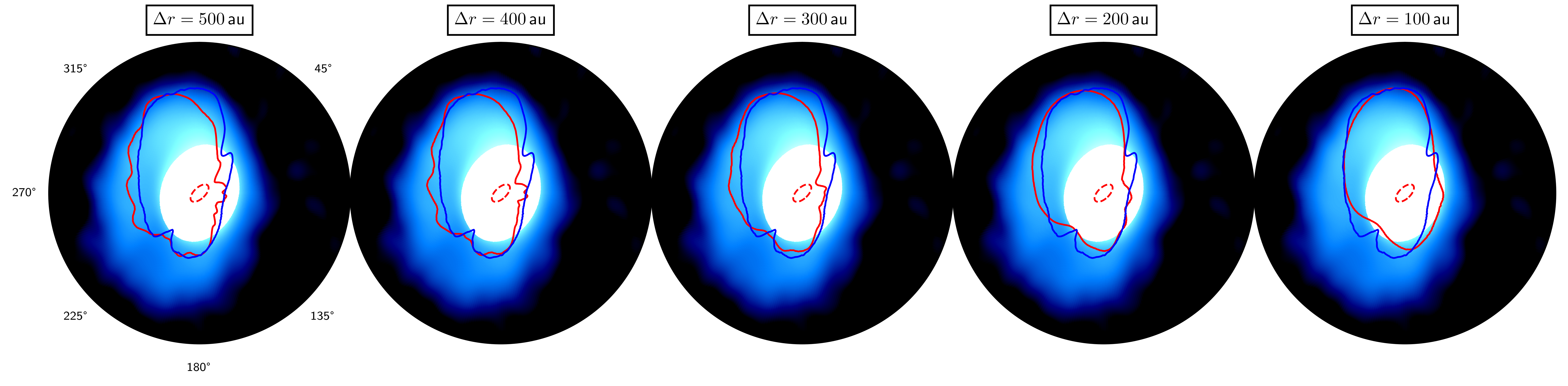}}
      \caption{Comparison of normalized azimuthal profiles (red curves) for five different radial integration distances $\Delta r$ based on run 5 at $150 \,$kyr (corresponding to the synthetic ALMA observation shown in Fig.~\ref{fig:alma_m154_400pc_wAzimuthalProfiles}). The dashed red ellipse in the center of each plot shows the underlying assumed inner disk, whose properties were chosen according to those of the approximate ideal inner disk. The blue curves show ideal azimuthal profiles. Each red curve corresponds to the obtained normalized azimuthal profile of each plot, assuming the $\Delta r$ value above the plot. 
              }
         \label{fig:different_az_profiles.pdf}
   \end{figure*}

Next, we compared different possible azimuthal profiles. Simply integrating the flux over a narrow azimuthal range in radial direction, as has been done for the ideal azimuthal profiles, did not suffice as it leads to a great abundance of minima in the profile. This occurs due to the fact, that the integration is performed over a region that is for the most part dominated by noise. We find that the quality of the azimuthal profile improves when restricting the radius, up to which the integration is performed. In particular, the best results were obtained when restricting the radial integration range to a constant length $\Delta r$, i.e., depending on the direction $\phi$, the integration was performed starting from the outer edge of the masked region up to a distance of $\Delta r$. However, since the corresponding integrated area $A_{\rm \phi}$ hence depends on the shape of the masked region, it is furthermore required to normalize the azimuthal profile for every direction according to $A_{\rm \phi}$. Based on a comparison of different values of $\Delta r$ for various models, configurations, and wavelengths (compare with Fig.~\ref{fig:different_az_profiles.pdf}), we find that for our synthetic observations $\Delta r=300\,$au often leads to the most reliable results and the most suitable azimuthal profiles for detecting shadows in disks. We note, that the best value for $\Delta r$ strongly depends on the quality of the observation and needs to be determined for each observation individually. 

\begin{figure*}
   \resizebox{\hsize}{!}
            {\includegraphics[width=\textwidth,clip]{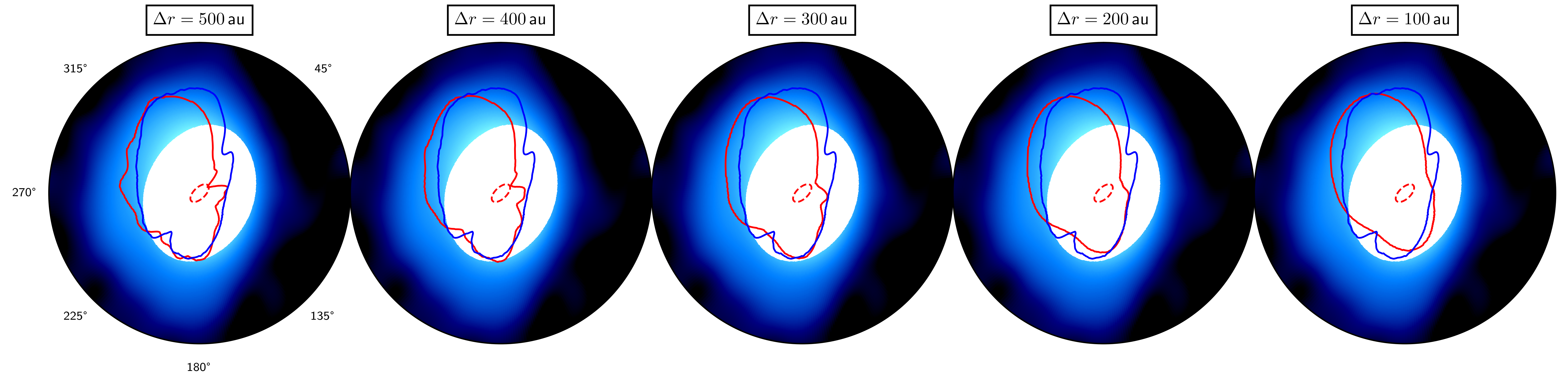}}
      \caption{Similar to Fig.~\ref{fig:different_az_profiles.pdf}, now for a synthetic band 6 ALMA observation of the same system at $1300\,\mu$m.
              }
         \label{fig:different_az_profiles_1300.pdf}
   \end{figure*}

Figure \ref{fig:different_az_profiles.pdf} exemplarily shows the resulting normalized azimuthal profiles for different values of $\Delta r$ ranging from $100$ to $500\,$au (red curves). These plots suggest that in the considered range, higher value of $\Delta r$ lead to deeper dips in the azimuthal profile at the cost of a greater number of dips, i.e., the structure of the azimuthal profile becomes more complex. On the contrary, low values of $\Delta r$ typically lead to more shallow dips and a much smoother shape of the azimuthal profile. In the case of $\Delta r=300\,$au, for instance, two almost diametrically opposed dips can be seen, that hint at the presence of an underlying shadow, while other regions in the azimuthal profile are for the most part smoother. We also find, that these two dips appear at a similar position as the dips that are present in the ideal map. However, there are also different other minima present in the azimuthal profile, for instance in the direction $\phi\approx135^\circ$, which are not present in the ideal azimuthal profile (blue curve). We note, that in a similar analysis of synthetic ALMA observations of the same system at $1300\,\mu$m (band 6), we altogether found very similar trends, however, the shadow feature appears to be even more spread out due to the increased beam size. The corresponding azimuthal profiles are shown in Fig.~\ref{fig:different_az_profiles_1300.pdf}. Although the existence of two almost diametrically opposed dips in the azimuthal profile gives a strong indication for the presence of a shadow, false detections cannot be fully ruled out using this method alone. Moreover, it is generally difficult to decide which of the most suitable inner disk guesses is leading to the best azimuthal profile, unless the approximate shape of the inner disk is already known due to observations performed for instance in the VIS/NIR wavelength range (see Fig.~\ref{fig:example_sphere_observation}). Generally, we find that the most suitable inner disk guesses with small values of $A_{\rm mask}$ lead to azimuthal profiles that are comparably smooth, while those with high $A_{\rm mask}$ values lead to azimuthal profiles with rather complex structures. Overall, this indicates that the most reliable strategy for detecting shadows in disks requires an analysis of these most suitable inner disk guesses and their corresponding azimuthal profiles and a search for a common shadow feature across those profiles, which would appear in the form of diametrically opposed dips. However, in the case of a high level of noise in the synthetic ALMA observations, as it is present in our simulations, it appears to be rather difficult to perform an unambiguous detection of shadows. Additionally, the size of the beam may often exceed the width of the shadow in the regions close to the inner edge of the outer disk, leading to a reduction of the contrast in brightness between the shadow feature and neighboring regions. Overall, applying the described method to real observations may provide evidence for the existence of shadows and justify an in-depth investigation of such a system regarding the possibility of a late infall event.

     \newpage

\FloatBarrier
\section{Supplementary material}

\FloatBarrier
\subsection{Azimuthal profiles}
\label{sec:app:traditional_azimuthal_profiles}
   \begin{figure}[!htb]
   \centering
   \includegraphics[width=\hsize]{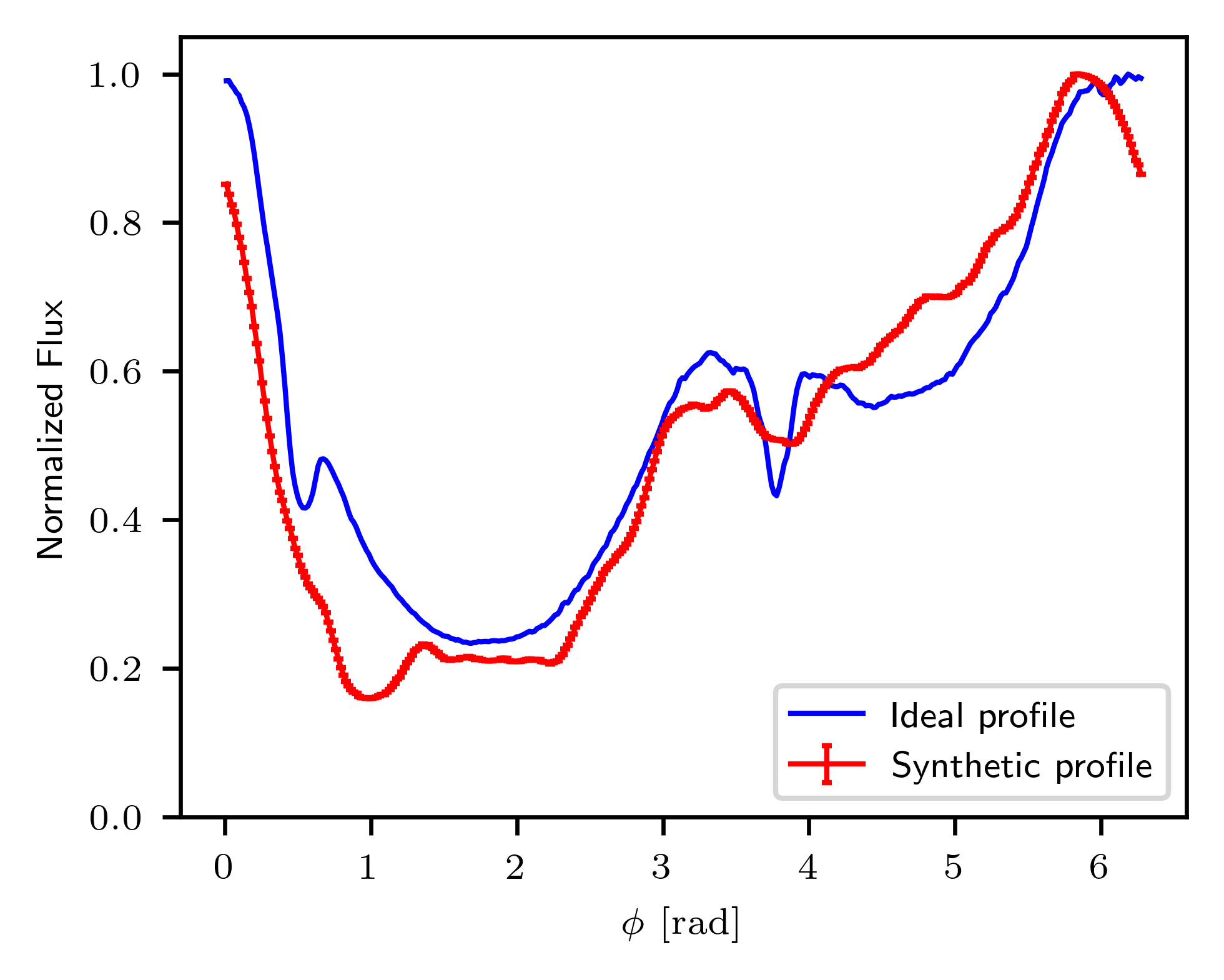}
      \caption{Normalized azimuthal profiles for an ALMA observation at $850\,\mu$m. For details, see Sect.~\ref{sec:app:traditional_azimuthal_profiles}.
              }
         \label{fig:traditional_azimuthal_profiles}
   \end{figure}

Figure \ref{fig:traditional_azimuthal_profiles} shows normalized azimuthal profiles for the ideal (red) and the synthetic (blue) ALMA observation at $850\,\mu$m using configuration C43-4 based on the simulation of run 5 at 150 kyr assuming a distance of $400\,$pc. These profiles correspond to the profiles shown in Fig.~\ref{fig:alma_m154_400pc_wAzimuthalProfiles}. For details, see Sect.~\ref{sec:alma_larger_dist}.

\FloatBarrier
\subsection{Synthetic observations}
\label{sec:app:more_syn_observations}
\begin{figure*}
   \centering
   \includegraphics[width=\hsize]{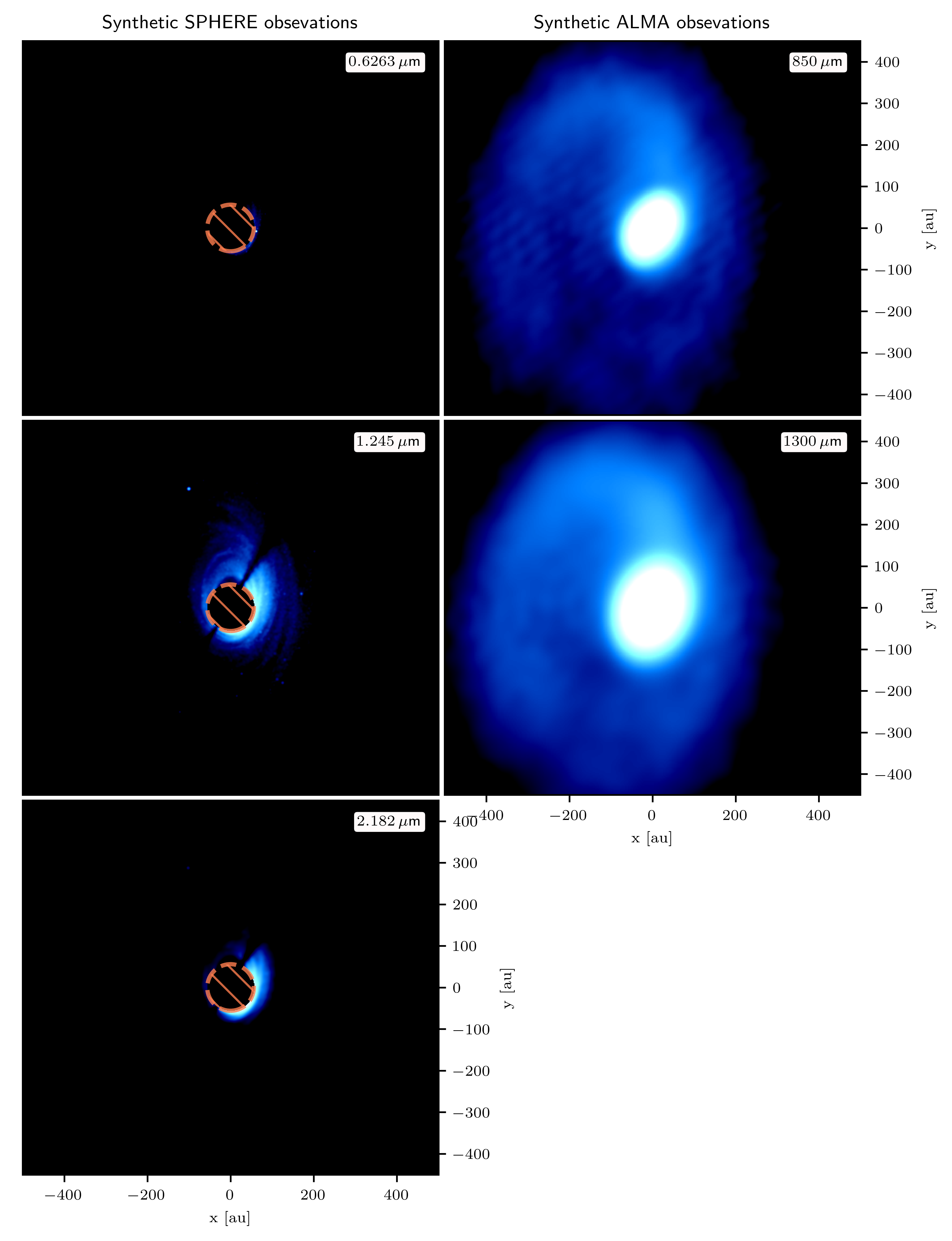}
      \caption{Collection of different synthetic SPHERE (left column) and ALMA observations (right column). For details, see Sect.~\ref{sec:app:more_syn_observations}.
              }
         \label{fig:alma_multi_bands}
   \end{figure*}

Figure \ref{fig:alma_multi_bands} shows synthetic SPHERE observations (left column) and synthetic ALMA observations (right column) for different bands. In particular, the synthetic ALMA observations are based on run 5 at 150 kyr and assume the usage of configuration C43-4 (upper right plot: rms=$3.32\cdot 10^{-4}\,$Jy/Beam; middle right plot: rms=$1.66\cdot 10^{-4}\,$Jy/Beam). Synthetic SPHERE observations are shown in the form of derived contrast maps based on the same run. For illustrative purposes, the hatched area of radius $400\,$mas in the centers of these plots is masked, and the color bar has been limited to the maximum range of detectable contrast values outside the masked region. Moreover, a distance of $140\,$pc is assumed for all five synthetic observations and the corresponding observing wavelength is displayed in the top right corner of each plot. 

\FloatBarrier
\subsection{Time evolution}
\label{sec:app:time_evolution}
   \begin{figure}[!htb]
   \centering
   \includegraphics[width=0.95\hsize]{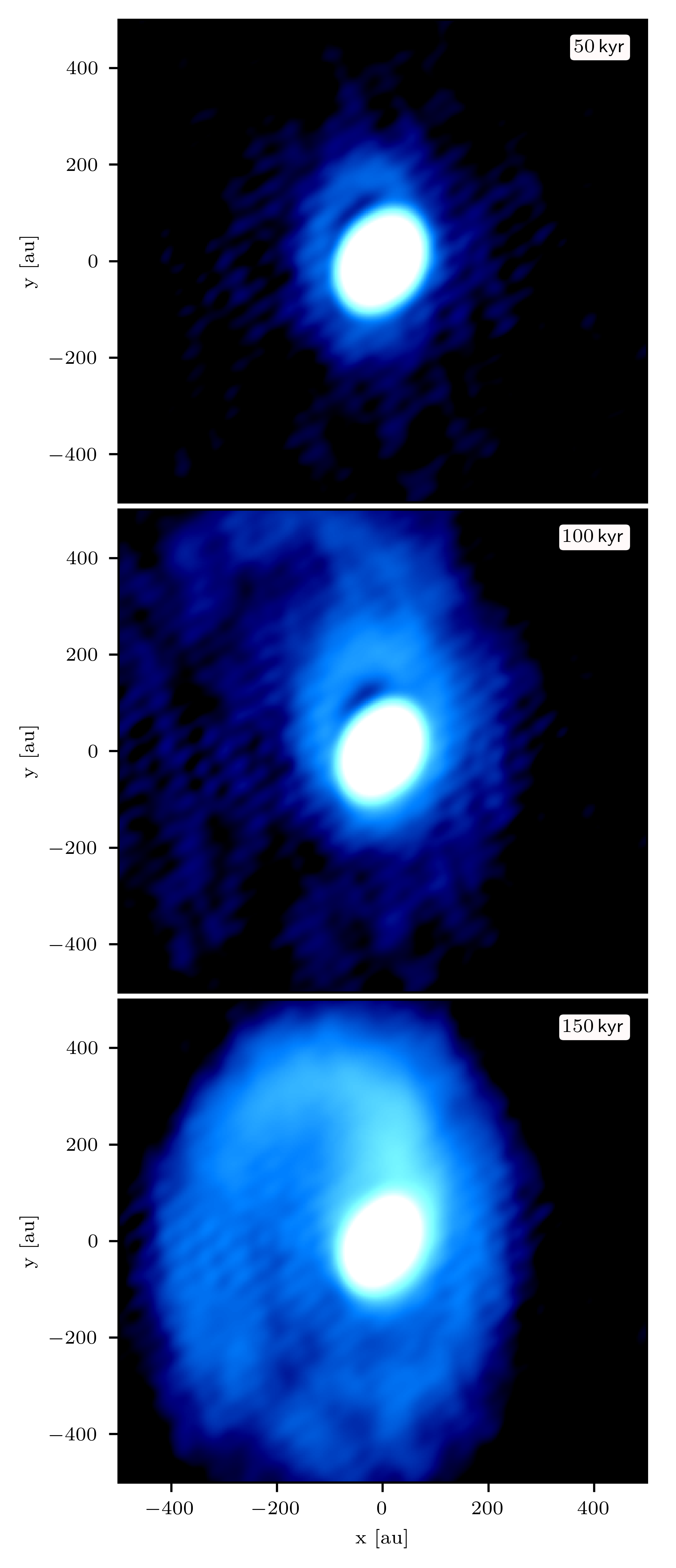}
      \caption{Synthetic ALMA observations at $850\,\mu$m. For details, see Sect.~\ref{sec:app:time_evolution}.}
         \label{fig:alma_m138_146_154}
   \end{figure}
Figure \ref{fig:alma_m138_146_154} shows synthetic ALMA observations at $850\,\mu$m using configuration C43-4 based on three different snapshots of run 5 assuming a distance of $140\,$pc: at 50\,kyr (upper plot; rms=$3.76\cdot 10^{-4}\,$Jy/Beam), 100\,kyr (middle plot; rms=$3.47\cdot 10^{-4}\,$Jy/Beam), and 150\,kyr (lower plot; rms=$3.32\cdot 10^{-4}\,$Jy/Beam).

\FloatBarrier
\subsection{Multi-configuration observations}
\label{sec:app:combination}
   \begin{figure}[!htb]
   \centering
   \includegraphics[width=\hsize]{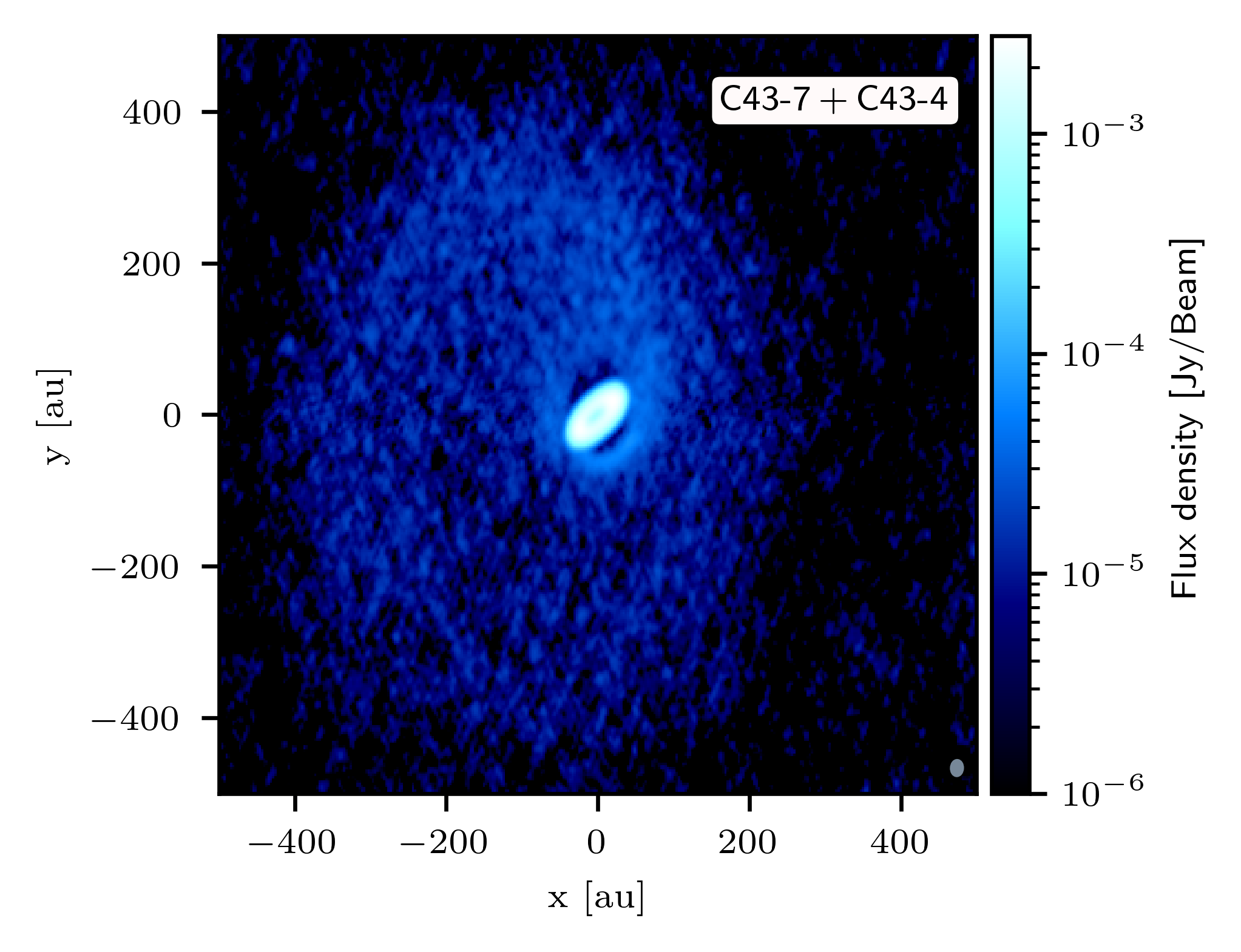}%rms 4.040577737272121e-05
      \caption{Synthetic ALMA observations at $850\,\mu$m. For details, see Sect.~\ref{sec:app:combination}.}
         \label{fig:combination_alma_C4plus7}
   \end{figure}

Figure \ref{fig:combination_alma_C4plus7} shows a synthetic ALMA observation at $850\,\mu$m using a combination of observations with configuration C43-7 and C43-4 (rms=$4.04\cdot 10^{-5}\,$Jy/Beam) based on run 5 at $150 \,$kyr assuming a distance of $140\,$pc. The corresponding beam (gray) is displayed in the lower right corner of the plot. In particular, the interferometric data obtained from simulated observation with configuration C43-7 (see Fig. \ref{fig:alma_m154_CASA_vs_CASAx10}) has been expanded by a data set obtained for configuration C43-4 in accordance with observatory-recommended practices by using a observation time ratio of {C43-7\,${:}$\,C43-4\,=\,1\,${:}$\,0.23}\footnote{See ALMA technical handbook, Table 7.5: \\\url{https://almascience.eso.org/documents-and-tools/cycle9/alma-technical-handbook}}. This combination provides the large angular scale of configuration C43-4 with the high resolution of configuration C43-7. For further details, see Sect.~\ref{sec:alma_sensitivity}.

\FloatBarrier
\subsection{Instrument sensitivity}
\label{sec:app:x10flux}
\begin{figure*}
   \centering
   \includegraphics[width=\hsize]{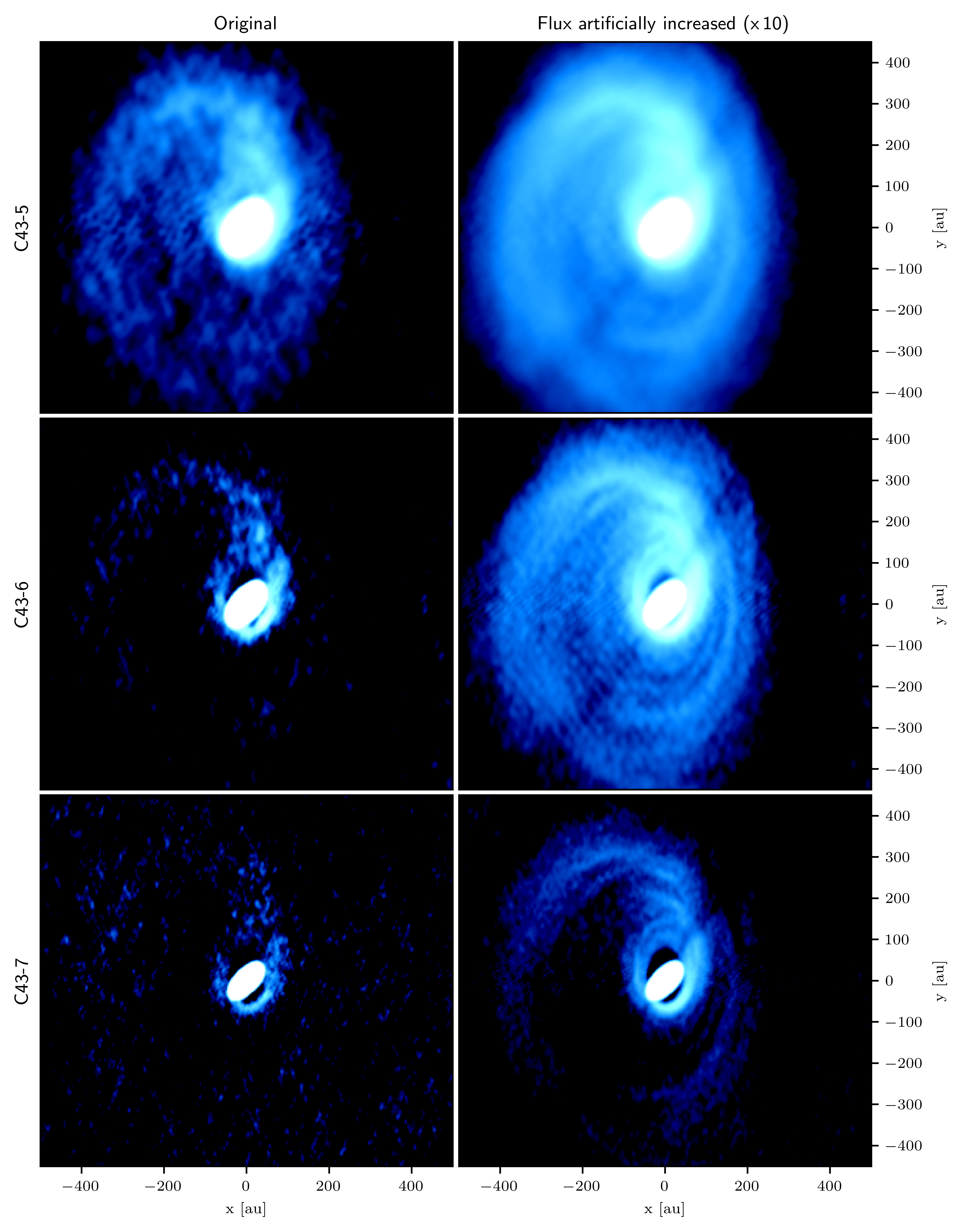}
      \caption{Synthetic ALMA observations at $850\,\mu$m. For details, see Sect.~\ref{sec:app:x10flux}.}
         \label{fig:alma_m154_CASA_vs_CASAx10}
   \end{figure*}

Figure \ref{fig:alma_m154_CASA_vs_CASAx10} shows synthetic ALMA observations at $850\,\mu$m using configurations C43-5 (upper row; left plot: rms=$1.62\cdot 10^{-4}\,$Jy/Beam; right plot: rms=$1.62\cdot 10^{-3}\,$Jy/Beam), C43-6 (middle row; left plot: rms=$6.60\cdot 10^{-5}\,$Jy/Beam; right plot: rms=$6.61\cdot 10^{-4}\,$Jy/Beam), and C43-7 (bottom row; left plot: rms=$3.09\cdot 10^{-5}\,$Jy/Beam; right plot: rms=$3.07\cdot 10^{-4}\,$Jy/Beam) based on run 5 at $150 \,$kyr assuming a distance of $140\,$pc. The left column shows results based on the unaltered ideal flux map, while the right column corresponds to the same ideal map with tenfold increased flux values. The corresponding beam (gray) is displayed in the lower right corner of each plot. For details, see Sect.~\ref{sec:alma_sensitivity}.

\FloatBarrier
\subsection{Column density}
\label{sec:app:column_density}
   \begin{figure}[!htb]
   \centering
   \includegraphics[width=\hsize]{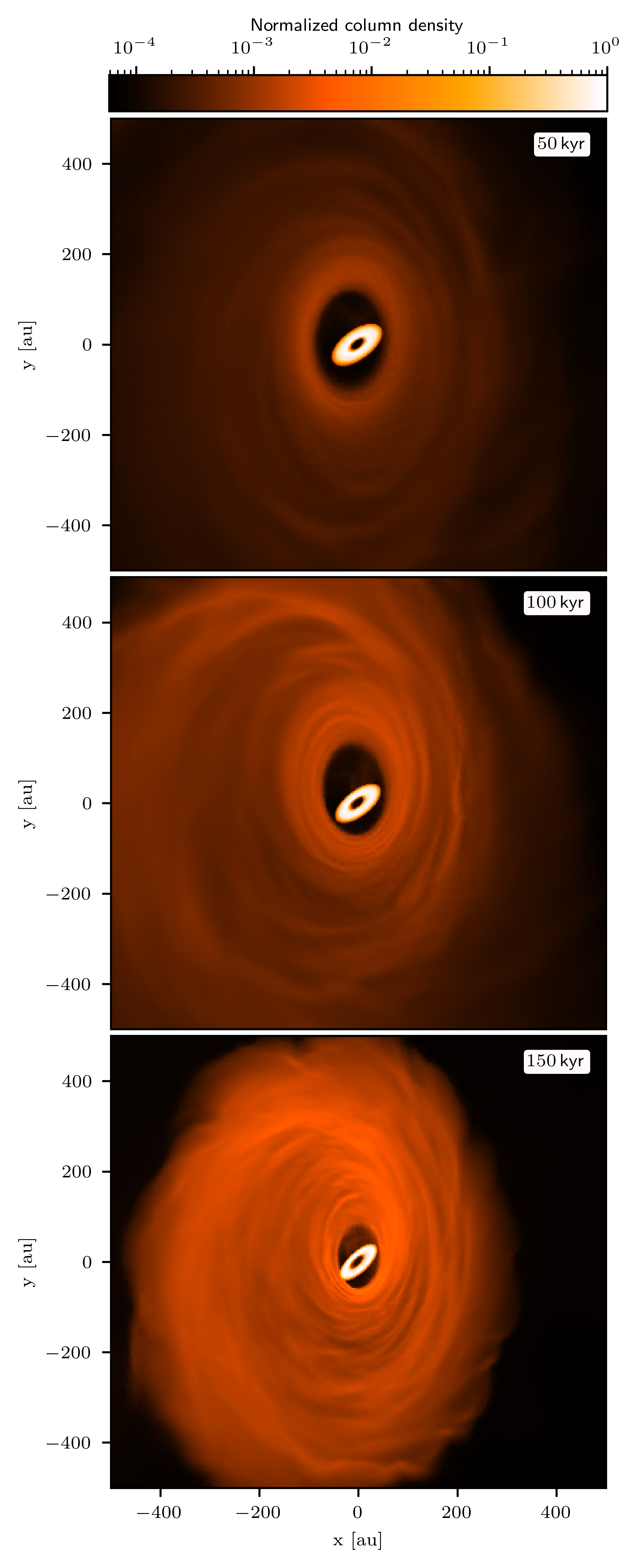}
      \caption{Corresponding column density of three snapshots of run 5 at 50\,kyr (upper plot), 100\,kyr (middle plot), and 150\,kyr (lower plot).}
         \label{fig:column_density}
   \end{figure}

\FloatBarrier
\subsection{Temperature distribution}
\label{sec:app:temp_dist}
   \begin{figure}[!htb]
   \centering
   \includegraphics[width=\hsize]{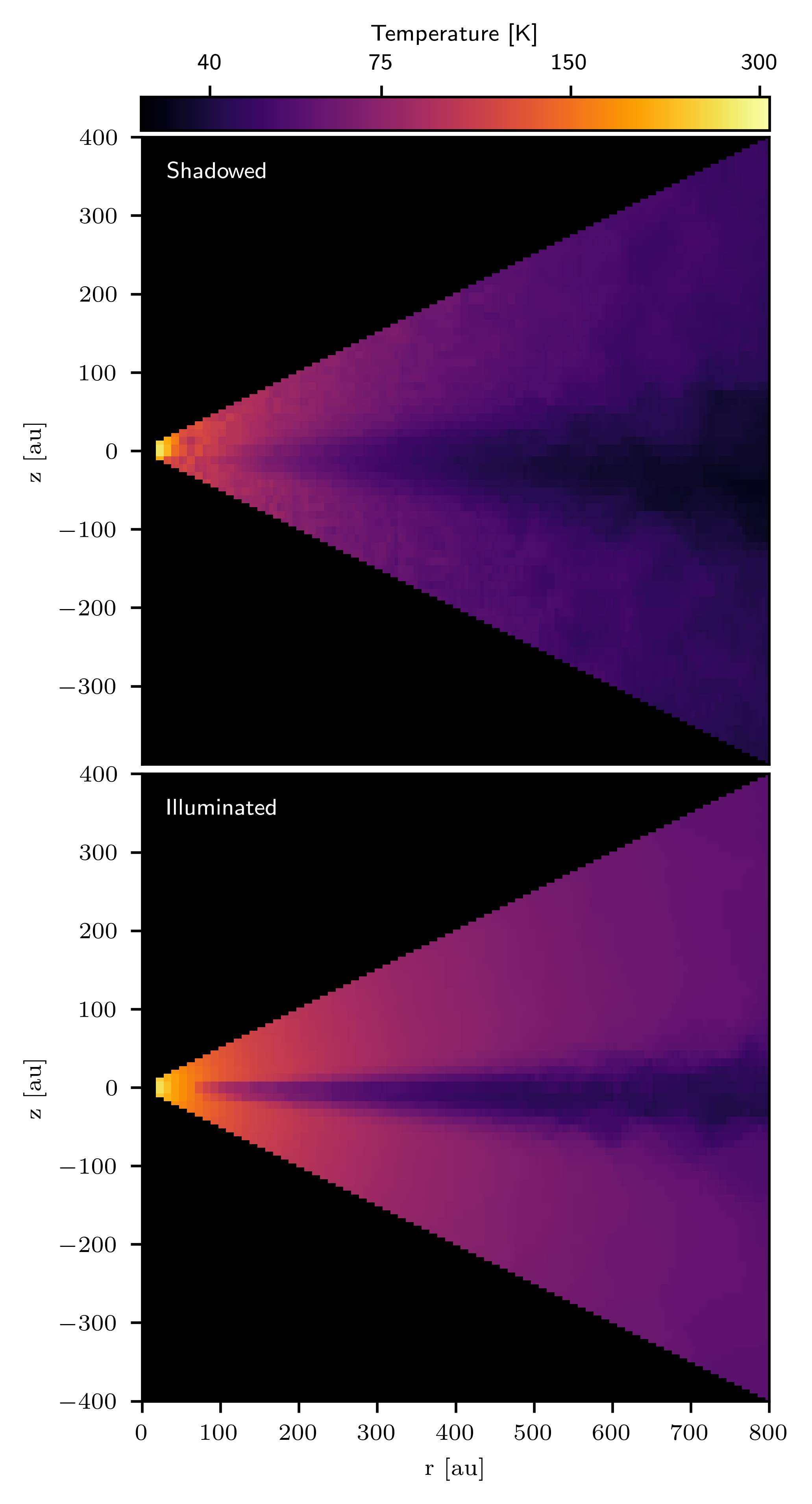}
      \caption{Temperature distributions corresponding to vertical cuts through the mid-plane of the outer disk. For details, see Sect.~\ref{sec:app:temp_dist}.}
         \label{fig:temp_dist}
   \end{figure}
Figure \ref{fig:temp_dist} shows an averaged temperature distribution, which is based on run 5 at $150 \,$kyr, corresponding to a vertical cut through the mid-planet of the outer disk which is either shadowed due to the inner disk (upper plot) or directly illuminated by the central star (lower plot). In particular, these results were obtained by averaging temperature distributions across an azimuthal range of $\Delta\phi = 0.1\pi$.

\end{appendix}
\end{document}